%% file: paper.tex
\documentclass[usenatbib]{mn2e}

\usepackage{amsmath}
\usepackage{natbib}
\usepackage{epsfig}
\usepackage{txfonts}
\usepackage{pict2e}
\usepackage{color}
\usepackage{booktabs}
\usepackage{multirow}

\input{Befehle}

\usepackage{hyperref}
\usepackage{grffile}
\usepackage{graphics}
\usepackage{booktabs}

\title[$\Lambda$CDM distortions]
{Which spectral distortions does $\Lambda$CDM actually predict?}

\author[Chluba]{Jens~Chluba$^{1}$\thanks{E-mail:Jens.Chluba@manchester.ac.uk}
\\
$^{1}$ Jodrell Bank Centre for Astrophysics, University of Manchester, Oxford Road, Manchester M13 9PL, UK
}

\begin{document}

\date{\vspace{-3.5mm}{Accepted 2016 April 18. Received 2016 March 12}}

\maketitle

\begin{abstract}
Ever refined cosmological measurements have established the $\Lambda$CDM concordance model, with the key cosmological parameters being determined to percent-level precision today. This allows us to make explicit {\it predictions} for the spectral distortions of the cosmic microwave background (CMB) created by various processes occurring in the early Universe. Here, we summarize all {\it guaranteed} CMB distortions and assess their total uncertainty within $\Lambda$CDM. We also compare simple methods for approximating them, highlighting some of the subtle aspects when it comes to interpreting future distortion measurements. Under simplified assumptions, we briefly study how well a {\it PIXIE}-like experiment may measure the main distortion parameters (i.e., $\mu$ and $y$). Next generation CMB spectrometers are expected to detect the distortion caused by reionization and structure formation at extremely high significance. They will also be able to constrain the small-scale power spectrum through the associated $\mu$-distortion, improving limits on running of the spectral index. Distortions from the recombination era, adiabatic cooling of matter relative to the CMB and dark matter annihilation require a higher sensitivity than {\it PIXIE} in its current design. The crucial next step is an improved modeling of foregrounds and instrumental aspects, as we briefly discuss here.
\end{abstract}

\begin{keywords}
Cosmology: cosmic microwave background -- theory -- observations
\end{keywords}

\section{Introduction}
\label{sec:Intro}
The standard $\Lambda$CDM cosmology has been shown to describe our Universe to extremely high accuracy \citep{WMAP_params, Planck2013params, Planck2015params}. This model is based upon a spatially flat, expanding Universe with dynamics governed by General Relativity and whose dominant constituents at late times are cold dark matter (CDM) and a cosmological constant ($\Lambda$). The primordial seeds of structures are furthermore Gaussian-distributed adiabatic fluctuations with an almost scale-invariant power spectrum thought to be created by inflation.

Today, we know the key cosmological parameters (e.g., the total, CDM and baryon densities, the CMB photon temperature, Hubble expansion rate, etc.) to percent-level precision or better \citep{Planck2015params}. Assuming standard Big Bang Nucleosynthesis (BBN) and a standard thermal history, we can furthermore derive precise values for the helium abundance, $\Yp$, and effective number of relativistic degrees of freedom, $N_{\rm eff}$ \citep[e.g.,][]{Steigman2007}. Also the physics of the recombination era, which defines the decoupling of photons and baryons around redshift $z\simeq 10^3$, is now believed to be well understood within $\Lambda$CDM \citep[e.g.,][]{Chluba2010b, Yacine2010c}.

Of the many cosmological data sets, measurements of the cosmic microwave background (CMB) temperature and polarization anisotropies, beyond doubt, have driven the development towards precision cosmology over the few past years. Today, cosmologists have exhausted practically all information about the primordial Universe contained in the CMB temperature power spectra. 
{\it WMAP} and {\it Planck} have also clearly seen the $E$-mode polarization signals \citep{Page2006, Planck2013powerspectra, Planck2015powerspectra}. Several sub-orbital and space-based experiments (e.g., {\it BICEP3, CLASS, SPTpol, ACTpol, SPIDER, PIPER, LiteBird, PIXIE, COrE+}) are now rushing to detect the primordial $B$-modes at large angular scales and to squeeze every last bit of information out of the $E$-mode signals, all to deliver the long-sought proof of inflation.

It is well known that CMB {\it spectral distortions} -- tiny departures of the average CMB energy spectrum from that of a perfect blackbody -- deliver a new independent probe of different processes occurring in the early Universe. The case for spectral distortions has been made several times and the physics of their formation is well understood \citep[for recent overview see,][]{Chluba2011therm, Sunyaev2013, Chluba2013fore, Tashiro2014, deZotti2015}. The purpose of this article is to provide an overview of all distortion signals created within $\Lambda$CDM (Fig.~\ref{fig:signals}), also assessing their total uncertainty. Although non-standard processes (i.e., decaying particles or evaporating primordial black holes) could cause additional interesting signals, the $\Lambda$CDM distortions define clear targets for designing future distortion experiments. 

Thus far, no all-sky distortion has been found \citep{Fixsen1996, Fixsen2009}, however, new experimental concepts, such as {\it PIXIE} \citep{Kogut2011PIXIE}, are being actively discussed and promise improvements of the measurements carried out with {\it COBE/FIRAS} by several orders of magnitude. It is thus time to ask what new information could be extracted from the CMB spectrum and how this could help refine our understanding of the Universe.

The case for spectral distortions as a new independent probe of inflation has also been made several times \citep[e.g.,][]{Hu1993, Chluba2012, Chluba2012inflaton, Dent2012, Pajer2012b, Khatri2013forecast, Chluba2013iso, Chluba2013fore, Clesse2014}, most recently by \citet{Cabass2016}, who emphasized that, given the constraints from {\it Planck}, an improvement in the sensitivity by a factor of $\simeq 3$ over {\it PIXIE} guarantees either a detection of $\mu$ or of negative running ($\gtrsim 95\%$ c.l.).
Here we add a few aspects to the discussion related to the interpretation of future distortion measurements carried out with an instrument similar to {\it PIXIE}. 
For real distortion parameter estimation, one has to simultaneously determine the average CMB temperature, $\mu$, $y$ and residual ($r$-type) distortion parameters, as well as several foreground parameters from measurements in different spectral bands \citep{Chluba2013PCA}. 
In this case, estimates for $\mu$ and $y$ based on simple scattering physics arguments (Sect.~\ref{sec:estimates}) underestimate the experimentally recovered ($\leftrightarrow$ measured) parameters, as we illustrate here.

We also briefly illustrate how well a {\it PIXIE}-like experiment may be able to constrain power spectrum parameters through the associated $\mu$-distortion when combined with existing constraints from {\it Planck} \citep{Planck2015params}. We find that an experiment with $\simeq 3.4$ times the sensitivity of {\it PIXIE} in its current design \citep{Kogut2011PIXIE} could allow tightening the constraint on the running of the spectral index by $\simeq 40\%-50\%$ when combined with existing data. This would also deliver an $\simeq 5\sigma$ detection of the $\mu$-distortion from CMB distortions alone. An $\simeq 10$ times enhanced sensitivity over {\it PIXIE} would furthermore allow a marginal detection of the first residual distortion parameter, which could be crucial when it comes to distinguishing different sources of distortions.

These forecasts are very idealized, assuming that the effective channel sensitivity already includes the penalty paid for foreground separation. Clearly, a more detailed foreground modeling for the monopole is required to demonstrate the full potential of future spectroscopic CMB missions, as we briefly discuss in Sect.~\ref{sec:foregrounds}. However, we argue that a combination of different data sets and exploitation of the many spectral channels of {\it PIXIE} will hopefully put us into the position to tackle this big challenge in the future.

\vspace{-3mm}
\section{Spectral distortions within $\Lambda$CDM}
\label{sec:distortions_LCDM}
Several exhaustive overviews on various spectral distortion scenarios exist \citep{Chluba2011therm, Sunyaev2013, Chluba2013fore, Tashiro2014, deZotti2015}, covering both standard and non-standard processes. Here we only focus on sources of distortions in the standard $\Lambda$CDM cosmology. For the numbers given in the text, we use the best-fitting parameters from {\it Planck} for the TT,TE,EE + lowP dataset \citep{Planck2015params}. Specifically, we use a flat model with $T_0=2.726\,\Kel$, $h=0.6727$, $\Omega_{\rm c}h^2=0.1198$, $\Omega_{\rm b}h^2=0.02225$, $\Yp=0.2467$ and $N_{\rm eff}=3.046$, with their standard meaning \citep{Planck2015params}.

\subsection{Reionization and structure formation}
\label{sect:reion}
The first sources of radiation during reionization \citep{Hu1994pert}, supernova feedback \citep{Oh2003} and structure formation shocks \citep{Sunyaev1972b, Cen1999, Refregier2000, Miniati2000} heat the intergalactic medium at low redshifts ($z\lesssim 10$), leading to a partial up-scattering of CMB photons, causing a Compton $y$-distortion \citep{Zeldovich1969}. 
Although this is the {\it largest} expected average distortion of the CMB caused within $\Lambda$CDM, its amplitude is quite uncertain and depends on the detailed structure and temperature of the medium, as well as scaling relations (e.g., between halo mass and temperature).
Several estimates for this contribution were obtained, yielding values for the total $y$-parameter at the level $y\simeq \pot{\rm few}{-6}$ \citep{Refregier2000, Zhang2004, Hill2015, Dolag2015, deZotti2015}. 

\begin{figure*}
\centering 
\includegraphics[width=2.0\columnwidth]{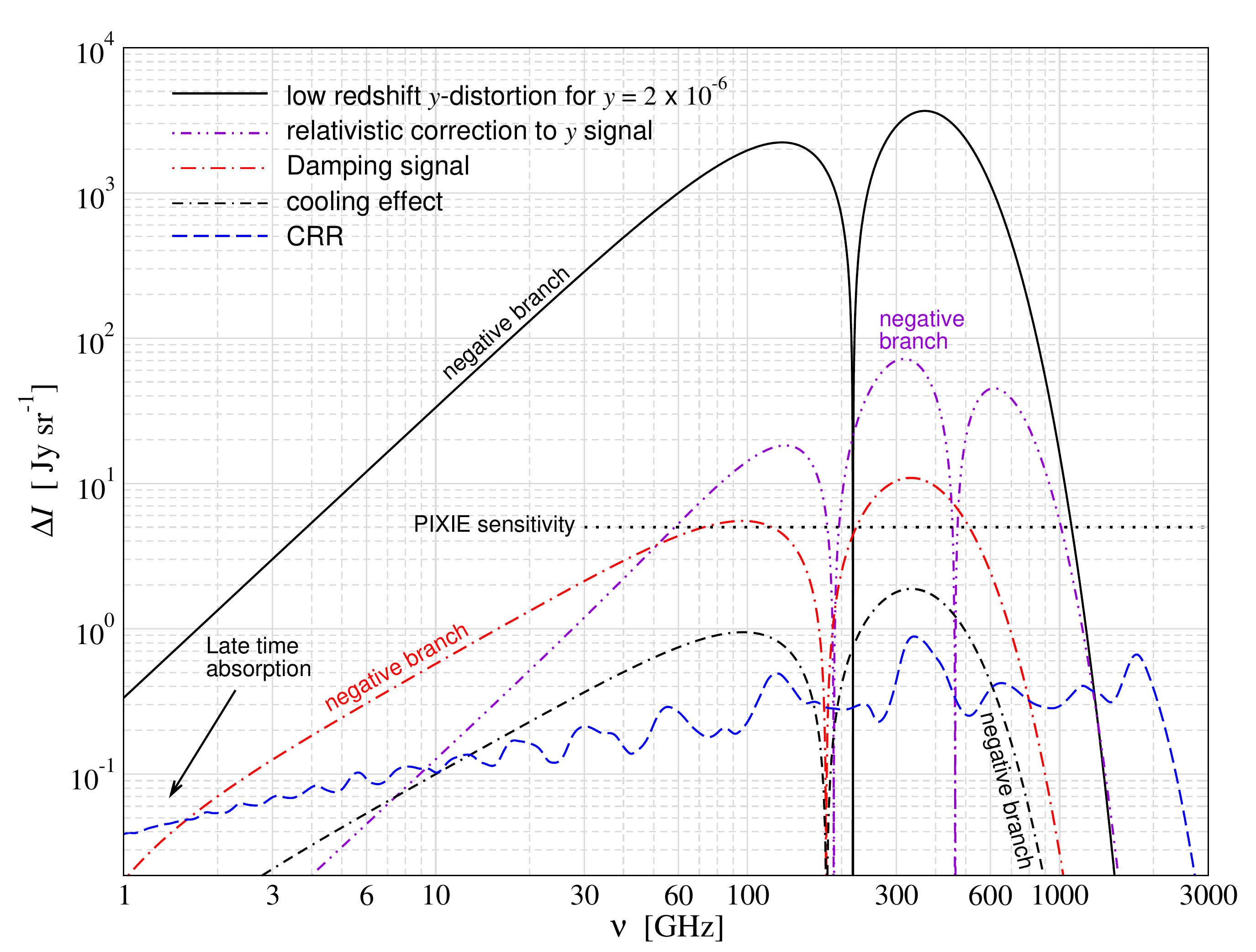}
\caption{Comparison of several CMB monopole distortion signals produced in the standard $\Lambda$CDM cosmology. 
The low-redshift distortion created by reionization and structure formation is close to a pure Compton-$y$ distortion with $y\simeq \pot{2}{-6}$. Contributions from the hot gas in low mass haloes give rise to a noticeable relativistic temperature correction, which is taken from \citet{Hill2015}. The damping and adiabatic cooling signals were explicitly computed using {\tt CosmoTherm} \citep{Chluba2011therm}. The cosmological recombination radiation (CRR) was obtained with
{\tt CosmoSpec} \citep{Chluba2016}. The estimated sensitivity ($\Delta I_\nu \approx 5\, {\rm Jy/sr}$) of PIXIE is shown for comparison (dotted line). The templates will be made available at {\tt www.Chluba.de/CosmoTherm}.}
\label{fig:signals}
\end{figure*}

Following  \citet{Hill2015}, we will use a fiducial value of $y_{\rm re}=\pot{2}{-6}$. This is dominated by the low mass end of the halo function and the signal should be detectable with {\it PIXIE} at more than $10^3\,\sigma$. At this enormous significance, small corrections due to the high temperature ($k\Te \simeq 1\,\keV$) of the gas become noticeable \citep{Hill2015}. The relativistic correction can be computed using the temperature moment method of {\tt SZpack} \citep{Chluba2012SZpack, Chluba2012moments} and it differs from the distortions produced in the early Universe. This correction should be detectable with {\it PIXIE} at $\simeq 30\,\sigma$ \citep{Hill2015} and could teach us about the average temperature of the intergalactic medium, promising a way to solve the missing baryon problem. Both distortion signals are illustrated in Fig.~\ref{fig:signals}.

\vspace{-0mm}
\subsection{Damping of primordial small-scale perturbations}
\label{sec:damp}
The damping of small-scale fluctuations of the CMB temperature set up by inflation at wavelength $\lambda<1\,\Mpc$  causes another inevitable distortion of the CMB spectrum \citep{Sunyaev1970diss, Daly1991, Barrow1991, Hu1994, Hu1994isocurv}. While the idea behind this mechanism is quite simple, it was only recently rigorously described \citep{Chluba2012}, allowing us to perform detailed computations of the associated distortion signal for different early universe models \citep{Chluba2012, Chluba2012inflaton, Dent2012, Pajer2012b, Khatri2013forecast, Chluba2013iso, Chluba2013fore, Clesse2014, Cabass2016}. The distortion is sensitive to the amplitude and shape of the power spectrum at small scales (wavenumbers $1\,\Mpc^{-1}\lesssim k \lesssim \pot{2}{4}\,\Mpc^{-1}$) and thus provides a promising new way to constrain inflation.

For a given initial power spectrum of perturbations, the effective heating rate in general has to be computed numerically. However, at high redshifts the tight coupling approximation can be used to simplify the calculation. An excellent approximation for the effective heating rate is obtained with\footnote{Here, we define the heating rate such that $\int_z^\infty \frac{\id (Q/\rho_\gamma)}{\id z}\id z>0$.} \citep{Chluba2012, Chluba2013iso}
\begin{align}
\label{eq:adiabatic_damping}
\frac{\id (Q/\rho_\gamma)}{\id z}&\approx 4 A^2 \partial_z \kD^{-2} \int^\infty_{k_{\rm min}} \frac{k^4\id k}{2\pi^2} P_\zeta(k)\,\expf{-2k^2/\kD^2},
\end{align}
where $P_\zeta(k)=2\pi^2\,A_{\rm s}\,k^{-3}\,(k/k_0)^{\nS-1 + \frac{1}{2}\,\nrun \ln(k/k_0)}$ defines the usual curvature power spectrum of scalar perturbations and $\kD$ is the photon damping scale \citep{Weinberg1971, Kaiser1983}, which scales as $\kD\approx \pot{4.048}{-6}\,(1 + z)^{3/2} \Mpc^{-1}$ early on. For adiabatic modes, we have a heating efficiency $A^2\approx (1+4R_\nu/15)^{-2}\approx 0.813$, where $R_\nu\approx 0.409$ for $N_{\rm eff}=3.046$. The $k$-space integral is truncated at $k_{\rm min}\approx 0.12\,\Mpc^{-1}$, which reproduces the full heating rate across the recombination era quite well \citep{Chluba2013fore}. With this we can directly compute the associated distortion using {\tt CosmoTherm} \citep{Chluba2011therm}. The various isocurvature perturbations can be treated in a similar manner \citep[e.g.,][]{Chluba2013iso}; however, in the standard inflation model these should be small. Tensor perturbations also contribute to the dissipation process; however, the associated heating rate is orders of magnitudes lower than for adiabatic modes even for very blue tensor power spectra and thus can be neglected \citep{Ota2014, Chluba2015}.

For $A_{\rm s}=\pot{2.207}{-9}$, $\nS=0.9645$ and $\nrun=0$ \citep{Planck2015params}, we present the result in Fig.~\ref{fig:signals}. The adiabatic cooling distortion (see Sect.~\ref{sec:ad_cool}) was simultaneously included. The signal is uncertain to within $\simeq 10\%$ in $\Lambda$CDM (Sect~\ref{sec:results}). The distortion lies between a $\mu$- and $y$-distortion and is close to the detection limit of {\it PIXIE}. 
As we will see below (Sect.~\ref{sec:forecasts}), with the current design the $\mu$-distortion part of the signal should be seen at the level of $\simeq 1.5\sigma$, which is in good agreement with earlier analysis \citep{Chluba2012, Chluba2013PCA}. A clear $5\sigma$ detection of this signal should be possible with $\simeq 3.4$ times higher sensitivity (Sect.~\ref{sec:forecasts}).
We will discuss various approximations for the damping signal below (Sect.~\ref{sec:Methods}), but simply performing a fit using $\mu$, $y$ and temperature shift \citep[see][for explicit definitions of these spectral shapes]{Chluba2013PCA}, $\Delta=\Delta T/T_0$, we find $\mu_{\rm fit}\approx \pot{1.984}{-8}$, $y_{\rm fit}\approx\pot{3.554}{-9}$ and $\Delta_{\rm fit}\approx \pot{-0.586}{-9}$ with a non-vanishing residual at the level of $20\%-30\%$.

\vspace{-1mm}
\subsection{Adiabatic cooling for baryons}
\label{sec:ad_cool}
The adiabatic cooling of ordinary matter continuously extracts energy from the CMB photon bath by Compton scattering leading to another small but guaranteed distortion that directly depends on the baryon density and helium abundance. The distortion is characterized by {\it negative} $\mu$- and $y$-parameters at the level of $\simeq \pot{\rm few}{-9}$ \citep{Chluba2005, Chluba2011therm, Khatri2011BE}. The effective energy extraction history is given by
\begin{align}
\label{eq:adiabatic_cooling}
\frac{\id (Q/\rho_\gamma)}{\id z}&=-\frac{3}{2}\,\frac{N_{\rm tot} k\Tg}{\rho_\gamma (1+z)} 
\nonumber
\\
&\approx -\frac{\pot{5.71}{-10}}{(1+z)}\,
\!\left[\frac{(1-\Yp)}{0.7533}\right]
\!\left[\frac{\Omega_{\rm b}h^2}{0.02225}\right]
\nonumber
\\
&\qquad\qquad\qquad\qquad\times\left[\frac{(1+f_{\rm He}+X_{\rm e})}{2.246}\right]\left[\frac{T_0}{2.726\,\Kel}\right]^{-3}
\end{align}
where $N_{\rm tot}=N_{\rm H}(1+f_{\rm He}+X_{\rm e})$ is the number density of all thermally coupled baryons and electrons; $N_{\rm H}\approx \pot{1.881}{-6}\,(1+z)^3\,\cm^{-3}$ is the number density of hydrogen nuclei; $f_{\rm He}\approx \Yp/4(1-\Yp)\approx 0.0819$ and $X_{\rm e}=\Ne/N_{\rm H}$ is the free electron fraction, which can be computed accurately with {\tt CosmoRec} \citep{Chluba2010b}. For {\it Planck} 2015 parameters, the signal is shown in Fig.~\ref{fig:signals}. It is uncertain at the $\simeq 1\%$ level in $\Lambda$CDM (Sect~\ref{sec:results}) and cancels part of the damping signal; however, it is roughly one order of magnitude weaker and cannot be separated at the currently expected level of sensitivity of next generation CMB spectrometers.

\subsection{The cosmological recombination radiation}
The cosmological recombination process is associated with the emission of photons in free-bound and bound-bound transitions of hydrogen and helium \citep{Zeldovich68, Peebles68, Dubrovich1975}. This causes a small distortion of the CMB and the redshifted recombination photons should be visible as the cosmological recombination radiation (CRR), a tiny spectral distortion ($\simeq$ nK-$\mu$K level) present at mm to dm wavelength \citep[for overview see][]{Sunyaev2009}. The amplitude of the CRR depends directly on the number density of baryons in the Universe. The helium abundance furthermore affects the detailed shape of the recombination lines. Finally, the line positions and widths depend on when and how fast the Universe recombined. The CRR thus provides an independent way to constrain cosmological parameters and  map the recombination history \citep{Chluba2008T0}. 

Several computations of this CRR have been carried out in the past \citep{Rybicki93, DubroVlad95, Dubrovich1997, Kholu2005, Jose2006, Chluba2006b, Chluba2007, Jose2008, Chluba2009c, Chluba2010}. These calculations were very time-consuming, taking a few days of supercomputer time for one cosmology \citep[e.g.,][]{Chluba2007, Chluba2010}. This big computational challenge was recently overcome \citep{Yacine2013RecSpec, Chluba2016}, today allowing us to compute the CRR in about 15 seconds on a standard laptop using {\tt CosmoSpec}\footnote{{\tt www.Chluba.de/CosmoSpec}} \citep{Chluba2016}. 
The {\it fingerprint} from the recombination era shows several distinct spectral features that encode valuable information about the recombination process (Fig.~\ref{fig:signals}). Many subtle radiative transfer and atomic physics processes \citep[e.g.,][]{Chluba2007, Chluba2009c, Chluba2010b, Yacine2010c} are included by {\tt CosmoSpec}, yielding the most detailed and accurate predictions of the CRR in the standard $\Lambda$CDM model to date. In $\Lambda$CDM, the CRR is uncertain at the level of a few percent, with the error being dominated by atomic physics \citep[see][]{Chluba2016}.

The CRR is currently roughly $\simeq 6$ times below the estimated detection limit of {\it PIXIE} (cf. Fig.~\ref{fig:signals}) and a detection from space will require several times higher sensitivity \citep{Vince2015}, which in the future could be achieved by experimental concepts similar to {\it PRISM} \citep{PRISM2013WPII} or {\it Millimetron} \citep{Smirnov2012}. At low frequencies ($1\,\GHz\lesssim \nu\lesssim 10\,\GHz$), the significant spectral variability of the CRR may also allow us to detect it from the ground \citep{Mayuri2015}.

\subsection{Superposition of blackbodies}
\label{sec:sup}
It is well-known that the superposition of blackbodies with different temperatures no longer is a blackbody but exhibits a $y$-type spectral distortion \citep{Zeldovich1972, Chluba2004, Stebbins2007}. It is precisely this effect that leads to the distortion caused by Silk-damping \citep[e.g.,][]{Chluba2012}. To second order in the temperature fluctuations $\Delta_T=\Delta T/\bar{T}\ll 1$, the effective $y$-parameter is given by \citep[cf.,][]{Chluba2004}
\begin{align}
\label{eq:y-sup}
y&=\frac{1}{2}\left<\Delta^2_T\right>,
\end{align}
where $\bar{T}=\left<T\right>$ and the average can be related to any blackbody intensity mixing process. This can be i) Thomson scattering, ii) weighted averages of CMB {\it intensity} maps (e.g., due to spherical harmonic decomposition) or iii) inevitable averaging inside the instrumental beam \citep{Chluba2004}. As mentioned above, i) occurs in the early Universe and also during reionization \citep[e.g.,][]{Hu1994pert, Chluba2012} while ii) can occur as an {\it artefact} of the standard analysis of CMB intensity/antenna temperature maps, which can in principle be avoided by consistently converting to thermodynamic temperature at second order in $\Delta_T$.

For the standard CMB temperature anisotropies, the beam averaging effect is tiny for angular resolution typical for today's CMB imaging experiments ({\it Planck}, SPT, ACT, etc.) and should be completely negligible \citep{Chluba2004}. For a {\it PIXIE}-type experiment with beam $\simeq 2^\circ$, the effect should be limited to $y\lesssim \pot{{\rm few}}{-10}-10^{-9}$ \citep{Chluba2004}; however, since high-resolution CMB temperature maps are available, this unavoidable effect can be taken into account very accurately.

The largest distortion due to the superposition of blackbodies with different temperatures is caused by the presence of the CMB dipole, with $\beta=\varv/c\simeq \pot{(1.231\pm 0.003)}{-3}$ \citep{Hinshaw2009}. Averaging the CMB intensity\footnote{We emphasize again that this distortion is {\it not} produced if the thermodynamic temperature, computed to second order in the fluctuations, was used.} over the whole sky yields \citep{Chluba2004}
\begin{align}
\label{eq:y-sup_CMB_dipole}
y_{\rm d}&=\frac{\beta^2}{6}\approx \pot{(2.525 \pm 0.012)}{-7}.
\end{align}
with sky-averaged temperature dispersion $\left<[\beta \cos(\theta)]^2\right>=\beta^2/3$. The uncertainty, $\Delta y_0^{\rm d}\approx \pot{1.2}{-9}$, in the contribution of the CMB dipole to the average $y$-parameter 
is a few times larger than the $y$-parameter induced by multipoles $\ell>1$ (see below). 
%
%
One may also worry about higher order temperature terms from the dipole \citep{Chluba2004}, however, only even moments contribute, so that the next order correction averaged over the whole sky, $\left<[\beta \cos(\theta)]^4\right>\simeq \beta^4/5\simeq \pot{(4.59\pm0.04)}{-13}$, is negligible, although it could still contribute noticeably in the far Wien tail ($\nu\gtrsim 1\,{\rm THz}$). We mention that in the presence of a primordial dipole, we would in addition obtain a $y$-parameter contribution $y_0^{\rm pD}\simeq \beta \Delta_{10}/\sqrt{12\pi}$, which is caused by aberration and Doppler boosting \citep{Chluba2011ab} and could reach $\simeq 10^{-9}$. Here, $\Delta_{10}$ is the spherical harmonic coefficient of the primordial dipole along the direction of the CMB dipole.

Averaging the CMB intensity spectrum over the whole sky (after subtracting the CMB dipole spectrum), from the measured {\it Planck} temperature power spectrum \citep{Planck2015params} we find
\begin{align}
\label{eq:y-sup_CMB}
y_{\rm sup}&=\sum_{\ell=2} \frac{(2\ell+1)\,C_\ell}{8\pi}\approx\pot{8.23}{-10},
\end{align}
which is extremely close to an earlier estimate based on the theoretical CMB power spectrum \citep{Chluba2012}. The uncertainty in this derived value is dominated by large-angle foreground residuals but is estimated to be below $\simeq 1\%$. As this number is specific to our realization of the Universe and to our own location, it is {\it not} limited by cosmic variance. The CMB temperature is increased by $\Delta T\approx 4.49\,{\rm nK}$ due to the same effect.

We mention that the superposition of blackbodies of different temperatures has a slightly different physical effect than energy exchange through Compton scattering. In the latter case, energy is transferred from the electrons to the photons (assuming heating), such that photons are (partially) upscattered and afterwards {\it all} the energy is stored in the associated ($y$-type) distortion. For the mixing of blackbodies, {\it no} energy exchange between photons and electrons is required ($\leftrightarrow$ Thomson limit) and 2/3 of the energy stored by the original temperature fluctuations causes an increase for the average blackbody temperature. Thus, for a $y$-distortion created through the superposition of blackbodies one also finds an average temperature shift $\Delta T/\bar{T}=2 y_{\rm sup}$ \citep{Chluba2012, Chluba2015IJMPD, Inogamov2015}. For the effect of the dipole this implies $\Delta T_{\rm d}/T_0=\beta^2/3$ caused by the superposition. However, another correction, $\Delta T_{\rm D}/T_0\approx -\beta^2/2$, arises from the Lorentz boost, so that the total temperature shift is $\Delta T=\Delta T_{\rm d}+\Delta T_{\rm D}=-T_0\,\beta^2/6\approx -(0.688 \pm 0.003) \,\mu\Kel$ \citep{Chluba2004, Chluba2011ab}.

\vspace{-3mm}
\subsection{Dark matter annihilation}
Today, cold dark matter is a well-established constituent of our Universe \citep{WMAP_params, Planck2013params, Planck2015params}. However, the nature of dark matter is still unclear and many groups are trying to gather any new clue to help unravel this big puzzle \citep[e.g.,][]{Adriani2009, Galli2009, CDMS2010, Zavala2011, Huetsi2011, BSA11, Aslanyan2015}. Similarly, it is unclear how dark matter was produced, however, within $\Lambda$CDM, the WIMP scenario provides one viable solution \citep[e.g.,][]{Jungman1996, Bertone2005}. In this case, dark matter should annihilate at a low level throughout the history of the Universe and even today.

For specific dark matter models, the level of annihilation around the recombination epoch is tightly constrained with the CMB anisotropies \citep{Galli2009, Cirelli2009, Huetsi2009, Slatyer2009, Huetsi2011, Giesen2012, Diamanti2014, Planck2015params}. The annihilation of dark matter can cause changes in the ionization history around last scattering ($z\simeq 10^3$), which in turn can lead to changes of the CMB temperature and polarization anisotropies \citep{Chen2004, Padmanabhan2005, Zhang2006}. Albeit significant dependence on the interaction of the annihilation products with the primordial plasma \citep{Shull1985, Slatyer2009, Valdes2010, Galli2013, Slatyer2015}, the same process should lead to distortions of the CMB \citep{McDonald2001, Chluba2010a, Chluba2011therm}. 
The effective heating rate of the medium can be expressed as \citep[see also][]{Chluba2013fore}
\begin{align}
\label{eq:DM_annihil}
\frac{\id (Q/\rho_\gamma)}{\id z}&=f_{\rm ann}  \frac{N_{\rm H}(z)(1+z)^{2+\lambda}}{H(z)\,\rho_\gamma(z)} 
\end{align}
where $\lambda=0$ for s-wave annihilation. Here, $H$ denotes the Hubble factor and $\rho_\gamma$ the CMB photon energy density. The annihilation efficiency, $f_{\rm ann}$, captures all details related to the dark matter physics (e.g., annihilation cross section, mass, decay channels, etc.) and can be roughly taken as constant. 
For existing upper limits on $f_{\rm ann}$, the distortion is well below the detection limit of {\it PIXIE} \citep{Chluba2011therm, Chluba2013fore, Chluba2013PCA}. Using the latest constraints from {\it Planck}, we find the $\mu$-distortion to be $\mu\lesssim \pot{\rm few}{-10}-10^{-9}$ (see Sect.~\ref{sec:results}). For s-wave annihilation scenarios, this limit ought to be rather conservative, and it is hard to imagine a much larger effect. However, spectral distortion measurements are sensitive to {\it all} energy release at $z\lesssim \pot{2}{6}$ and not only limited to around last scattering. Thus, searches for this small distortion could deliver an important test of the WIMP paradigm should any signature of dark matter annihilation be found through another probe.
Possible coupling of WIMPs to the baryons or photons could further enhance the adiabatic cooling effect \citep{Yacine2015DM}, which could provide additional tests of the nature of dark matter especially for low dark matter masses.

\subsection{Anisotropic CMB distortions}
To close the discussion of different distortion signals, we briefly mention anisotropic ($\leftrightarrow$ {\it spectral-spatial}) CMB distortions.
Even in the standard $\Lambda$CDM cosmology, anisotropies in the spectrum of the CMB are expected. The largest source of anisotropies is due to the Sunyaev-Zeldovich effect caused by the hot plasma inside clusters of galaxies \citep{Zeldovich1969, Sunyaev1980, Birkinshaw1999, Carlstrom2002}. The $y$-distortion power spectrum has already been measured directly by {\it Planck} \citep{Planck2013ymap, Planck2015ymap} and encodes valuable information about the atmospheres of clusters \citep[e.g.,][]{Refregier2000, Komatsu2002, Diego2004, Battaglia2010, Shaw2010, Munshi2013, Dolag2015}. Similarly, the warm hot intergalactic medium contributes \citep{Zhang2004, Dolag2015}.

In the primordial Universe, anisotropies in the $\mu$ and $y$ distortions are expected to be tiny \citep[relative perturbations $\lesssim 10^{-4}$, e.g., see][]{Pitrou2010} unless strong spatial variations in the primordial heating mechanism are expected \citep{Chluba2012}. This could in principle be caused by non-Gaussianity of perturbations in the ultra-squeezed limit \citep{Pajer2012, Ganc2012, Biagetti2013, Razi2015}, however, this is beyond $\Lambda$CDM cosmology and will not be considered further.

Another guaranteed anisotropic signal is due to Rayleigh scattering of CMB photons in the Lyman-series resonances of hydrogen around the recombination era \citep{Yu2001, Lewis2013}. The signal is strongly frequency dependent, can be  modeled precisely and may be detectable with future CMB imagers (e.g., {\it COrE+}) or possibly {\it PIXIE} at large angular scales \citep{Lewis2013}. In a very similar manner, the resonant scattering of CMB photons by metals appearing in the dark ages \citep{Loeb2001, Zaldarriaga2002, Kaustuv2004, Carlos2007Pol} or scattering in the excited levels of hydrogen during recombination \citep{Jose2005, Carlos2007Pol} can lead to anisotropic distortions. To measure these signals, precise channel cross-calibration and foreground rejection is required. 

Due to our motion relative to the CMB rest frame, the spectrum of the CMB dipole should also be distorted simply because the CMB monopole has a distortion \citep{Danese1981, Balashev2015}. The signal associated with the large late-time $y$-distortion could be detectable with {\it PIXIE} at the level of a few $\sigma$ \citep{Balashev2015}. Since for these measurements no absolute calibration is required, this effect will allow us to check for systematics. In addition, the dipole spectrum can be used to constrain monopole foregrounds \citep{Balashev2015, deZotti2015}.

Finally, again due to the superposition of blackbodies (caused by the spherical harmonic expansion of the intensity map), the CMB quadrupole spectrum is also distorted, exhibiting a $y$-distortion related to our motion \citep{Kamionkowski2003, Chluba2004}. The associated effective $y$-parameter is $y_{\rm Q}=\beta^2/6 \approx \pot{(2.525 \pm 0.012)}{-7}$ and should be noticeable with {\it PIXIE} and future CMB imagers.

\section{Approximations for the distortion signals}
\label{sec:Methods}
The primordial distortion signals that are caused by early energy release can be precisely computed using {\tt CosmoTherm} \citep{Chluba2011therm}. However, for parameter estimation we will use Green's function method developed by \citet{Chluba2013Green}. The results for different scenarios will be compared with more approximate but very simple estimates summarized in the next section.

\subsection{Simple estimates for the $\mu$- and $y$-parameters}
\label{sec:estimates}
To compute estimates for the $\mu$- and $y$-parameters, several approximations have been discussed in the literature. Given the energy release history, $\id (Q/\rho_\gamma)/\id z$, they can all be compactly written as \citep[e.g.,][]{Chluba2013Green, Chluba2013PCA}
\bsub
\label{eq:Greens_approx_improved}
\begin{align}
y&=\frac{1}{4}\left.\frac{\Delta \rho_\gamma}{\rho_\gamma}\right|_y = \frac{1}{4}\int_0^\infty \mathcal{J}_y(z')\,\frac{\id (Q/\rho_\gamma)}{\id z'} \id z'
\\
\mu&=1.401\left.\frac{\Delta \rho_\gamma}{\rho_\gamma}\right|_\mu =1.401 \int_0^\infty \mathcal{J}_\mu(z') \frac{\id (Q/\rho_\gamma)}{\id z'} \id z'
\end{align}
\esub
where $\left.\Delta \rho_\gamma/\rho_\gamma\right|_y$ and $\left.\Delta \rho_\gamma/\rho_\gamma\right|_\mu$ denote the effective energy release in the $y$- and $\mu$-era, respectively. The individual distortion visibility functions, $\mathcal{J}_i(z)$, determine the differences between various existing approximations. The simplest approach assumes that the transition between $\mu$ and $y$ occurs sharply at $z=z_{\mu y}\simeq \pot{5}{4}$ and that no distortions are created at $z\gtrsim z_{\rm th}$, where $z_{\rm th}$ is the thermalization redshift, which is given by \citep{Burigana1991, Hu1993}
\begin{align}
\label{eq:z_th}
z_{\rm th}\approx
\pot{1.98}{6}\left[\frac{(1-\Yp/2)}{0.8767}\right]^{-2/5}
\!\left[\frac{\Omega_{\rm b}h^2}{0.02225}\right]^{-2/5}
\!\left[\frac{T_0}{2.726\,\Kel}\right]^{1/5}.
\end{align}
In this case, we have the simple approximation (`Method A')
\bsub
\begin{align}
\label{eq:M1}
\mathcal{J}_y(z)
&=
\begin{cases}
1 & \text{for}\; z_{\rm rec}\leq z \leq z_{\mu y}
\\
0 & \text{otherwise}
\end{cases}
\\
\mathcal{J}_\mu(z) 
&=
\begin{cases}
1 & \text{for}\; z_{\mu y}\leq z \leq z_{\rm th}
\\
0 & \text{otherwise}.
\end{cases}
\end{align}
\esub
For the estimates of $y$ from early energy release, we will not include any contributions from after recombination, $z\lesssim 10^3=z_{\rm rec}$. These contributions will be attributed to the reionization $y$-parameter.

The next improvement is achieved by taking into account that the thermalization efficiency does not abruptly vanish at $z\simeq z_{\rm th}$, but that even at $z>z_{\rm th}$ a small $\mu$-distortion is produced \citep{Sunyaev1970mu, Danese1982, Burigana1991, Hu1993}. With this we have (`Method B')
\bsub
\begin{align}
\label{eq:M1}
\mathcal{J}_y(z)
&=
\begin{cases}
1 & \text{for}\; z_{\rm rec}\leq z \leq z_{\mu y}
\\
0 & \text{otherwise}
\end{cases}
\\
\mathcal{J}_\mu(z) 
&=
\begin{cases}
\mathcal{J}_{\rm bb}(z) & \text{for}\; z_{\mu y}\leq z 
\\
0 & \text{otherwise}.
\end{cases}
\end{align}
\esub
where $\mathcal{J}_{\rm bb}(z)\approx \expf{-(z/z_{\rm th})^{5/2}}$ is the {\it distortion visibility function}.\footnote{Refined approximation for the distortion visibility function have been discussed \citep{Khatri2012b, Chluba2014}, but once higher accuracy is required it is easier to directly use the Green's function method, such that we do not go into more details here.}

\begin{figure}
\centering 
\includegraphics[width=1.05\columnwidth]{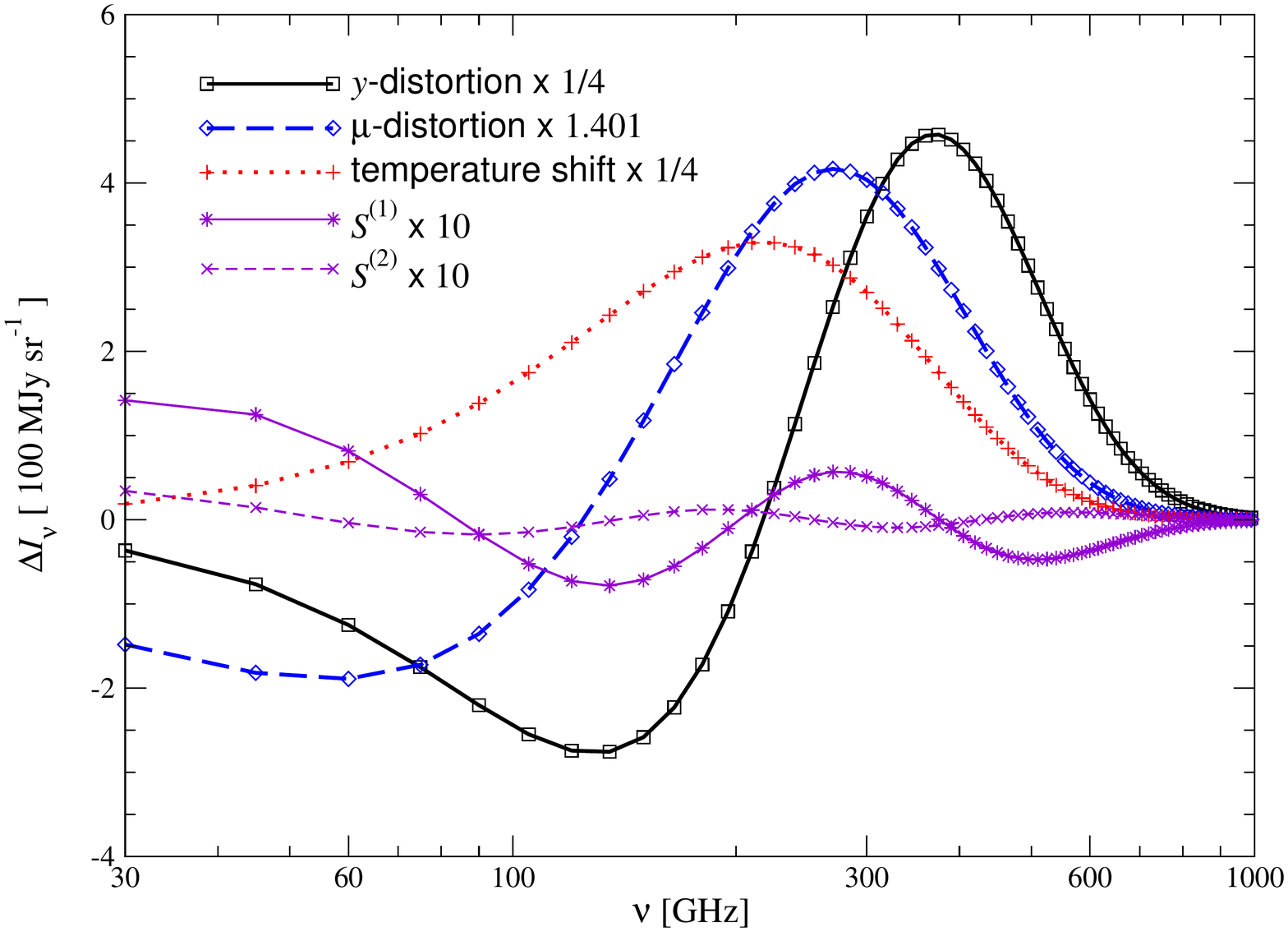}
\\
\includegraphics[width=1.05\columnwidth]{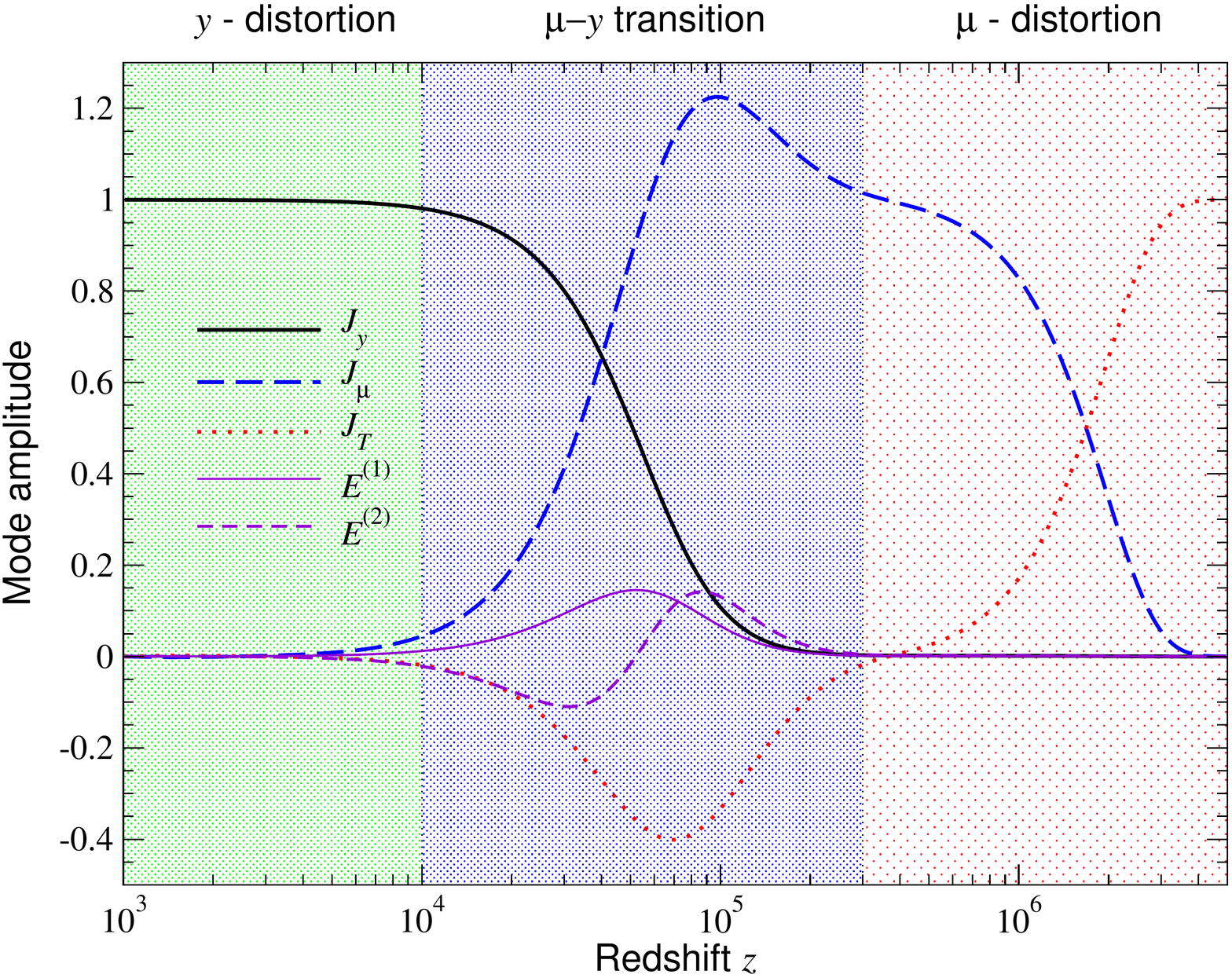}
\caption{Principal component decomposition for {\it PIXIE}-like setting ($\{\nu_{\rm min}, \nu_{\rm max}, \Delta\nu\}=\{30, 1000, 15\}\,\GHz$). -- Upper panel: first two residual distortion eigenmodes, $S^{(k)}$, in comparison with the spectral shapes of temperature shift, $\mu$ and $y$-distortions. We scaled the templates by convenient factors to make them comparable in amplitude. -- Lower panel: associated energy release eigenmodes, $E^{(k)}$, and visibilities, $J_i$, of temperature shift, $\mu$ and $y$-distortions. The figures were adapted from \citet{Chluba2013PCA}.}
\label{fig:Sk}
\label{fig:Ek}
\end{figure}

The next simple approximations also include the fact that the transition between $\mu$ and $y$ distortions is not abrupt at $z\simeq z_{\mu y}$. The distortion around this redshift is mostly given by a superposition of $\mu$ and $y$, with a smaller correction in form of the residual ($r$-type) distortion, which can be modeled numerically. By simply determining the best-fitting approximation to the distortion Green's function using only $\mu$ and $y$ one can write \citep{Chluba2013Green}
\bsub
\begin{align}
\label{eq:branching_approx_improved}
\mathcal{J}_y(z)
&\approx
\begin{cases}
\left(1+\left[\frac{1+z}{\pot{6}{4}}\right]^{2.58}\right)^{-1}
 & \text{for}\; z_{\rm rec}\leq z 
\\
0 & \text{otherwise}
\end{cases}
\\
\mathcal{J}_\mu(z) 
&\approx\mathcal{J}_{\rm bb}(z)\,\left[1-\exp\left(-\left[\frac{1+z}{\pot{5.8}{4}}\right]^{1.88}\right)\right].
\end{align}
\esub
We shall refer to this as `Method C' and only represents the exact proportions of $\mu$ and $y$ to $\simeq 10\%-20\%$ precision. To ensure full energy conservation (no leakage of energy to the $r$-distortion), instead one can use $\mathcal{J}_\mu(z)\approx [1- \mathcal{J}_y(z)]\,\mathcal{J}_{\rm bb}(z)$ (`Method D'). 

All the above expressions give slightly different results for the expected distortion $\mu$ and $y$-parameters. Below we will compare them with the more accurate distortion principal component decomposition \citep{Chluba2013PCA}, which optimizes the representation when simultaneously estimating $\mu$, $y$ and $\Delta=\Delta T/T_0$. At the same time, these approximations allow one to quickly estimate the expected distortion signals and their dependence on different parameters, which can be useful for order of magnitude work. We will see that a simple interpretation of the distortion in terms of $\mu$ and $y$ derived in this way differs slightly from what future measurements will recover (Sect.~\ref{sec:results}). Specifically, due to the uncertainty in the value of the CMB monopole, the projections of the distortion signals on to $\mu$ are underestimated by $\simeq 20\%-30\%$ (Table~\ref{tab:one}).

\begin{table*}
\centering
\caption{Comparison of distortion parameters for various methods and types of scenarios with cosmological parameters based on the {\it Planck} 2015 TT,TE,EE+lowP results \citep{Planck2015params}. 
The uncertainties for the predictions in the dissipation scenarios are dominated by the uncertainties in the power spectrum parameters.
Assuming standard BBN ($\Yp=0.2467$), for the adiabatic cooling distortion the small uncertainty is dominated by that of the baryon density. This contribution to the total distortion is also included in the values given for the dissipation scenarios. 
For the annihilation scenario, we assumed s-wave annihilation for a Mayorana dark matter particle using $p_{\rm ann}<\pot{4.1}{-28}\,\cm^3\,\sec^{-1}{\rm GeV}^{-1}$ \citep{Planck2015params} with $f_{\rm ann}=(\rho_{\rm c}^2 c^4 \Omega_{\rm cdm}^2/N_{\rm H, 0})\,p_{\rm ann}\approx \pot{8.5}{3}\,\eV\,\sec^{-1} p_{\rm ann}/[\cm^3\,\sec^{-1}{\rm GeV}^{-1}]$. All quoted error bars are for 68\% c.l. and the central values are medians.}
\begin{tabular}{cccccc}
\hline
\hline
 & Parameter
 & Dissipation I & Dissipation II
 & Adiabatic cooling
 & Annihilation (s-wave)  
\\
\hline
& $\ln(10^{10} A_{\rm s})$ 
& $3.094\pm0.034$ 
& $3.103 \pm 0.036$ 
& $-$ 
& $-$
\\  
& $\nS$ 
& $0.9645\pm0.0049$  
& $0.9639 \pm 0.0050$ 
& $-$
& $-$
\\
& $\nrun$ 
& $0$ 
& $\!\!\!\!-0.0057 \pm 0.0071$ 
& $-$
& $-$
\\
& $100\,\Omega_{\rm b} h^2$ 
& $2.225 \pm 0.016$ 
& $2.229 \pm 0.017$ 
& $2.225\pm0.016$
& $2.225\pm0.016$
\\
& $f_{\rm ann}$ 
& $-$ 
& $-$ 
& $-$
& $<\pot{3.5}{-24}\,\eV \, \sec^{-1} \,(\text{95\% c.l.})$
\\
\hline
\hline
\multirow{2}{*}{Method A}
& $y/10^{-9}$
& $3.67 ^{+ 0.17 } _{- 0.17 }$
& $3.53 ^{+ 0.25 } _{- 0.23 }$
& $-0.532^{+ 0.003} _{- 0.003}$
& $<0.091$
\\[2pt]
& $\mu/10^{-8}$
& $1.72 ^{+ 0.13 } _{- 0.12 }$
& $ 1.31 ^{+ 0.52 } _{- 0.38 } $
& $-0.296^{+ 0.002} _{- 0.002}$
& $<0.062$
\\[1pt]
\hline
\multirow{2}{*}{Method B}
& $y/10^{-9}$
& $3.67 ^{+ 0.18 } _{- 0.17 } $
& $3.54 ^{+ 0.25 } _{- 0.23 } $
& $-0.532^{+ 0.003} _{- 0.003}$
& $<0.091$
\\[2pt]
& $\mu/10^{-8}$
& $1.62 ^{+ 0.12 } _{- 0.11 }$
& $1.27 ^{+ 0.49 } _{- 0.36 } $
& $-0.277^{+ 0.002} _{- 0.002}$
& $<0.058$
\\[1pt]
\hline
\multirow{2}{*}{Method C}
& $y/10^{-9}$
& $3.83 ^{+ 0.19 } _{- 0.18 }$
& $3.66 ^{+ 0.28 } _{- 0.26 }$
& $-0.558^{+ 0.003} _{- 0.003}$
& $<0.097$
\\[2pt]
& $\mu/10^{-8}$
& $1.71 ^{+ 0.12 } _{- 0.12 }$
& $1.34 ^{+ 0.48 } _{- 0.36 }$
& $-0.290^{+ 0.002 } _{- 0.002 }$
& $<0.061$
\\[1pt]
\hline
\multirow{2}{*}{Method D}
& $y/10^{-9}$
& $3.83 ^{+ 0.19 } _{- 0.18 } $
& $3.66 ^{+ 0.29 } _{- 0.26 }$
& $-0.558^{+ 0.003} _{- 0.003}$
& $<0.097$
\\[2pt]
& $\mu/10^{-8}$
& $1.54 ^{+ 0.11 } _{- 0.11 }$
& $1.18 ^{+ 0.46 } _{- 0.33 }$
& $-0.263^{+ 0.001 } _{- 0.001 }$
& $<0.055$
\\[1pt]
\hline
\hline
\multirow{4}{*}{PCA}
& $y/10^{-9}$
& $3.63 ^{+ 0.17 } _{- 0.17 }$
& $3.49 ^{+ 0.26 } _{- 0.23 }$
& $-0.527^{+ 0.003} _{- 0.003}$
& $< 0.091$
\\[2pt]
& $\mu/10^{-8}$
& $2.00^{+ 0.14 } _{- 0.13 }$
& $1.59 ^{+ 0.54 } _{- 0.40 }$
& $-0.334^{+ 0.002} _{- 0.002}$
& $< 0.070$
\\[2pt]
& $\mu_1/10^{-8}$
& $3.81 ^{+ 0.22 } _{- 0.21 }$
& $3.39 ^{+ 0.58 } _{- 0.49 }$
& $-0.587 ^{+ 0.003} _{- 0.003 }$
& $\!\!\!< 0.12$
\\[2pt]
& $\mu_2/10^{-9}$
& $\!\!\!\!-1.19 ^{+ 0.22 } _{- 0.20}$
& $\!\!\!\!-2.79 ^{+ 2.05 } _{- 1.53 }$
& $-0.051 ^{+ 0.001} _{- 0.001 }$
& $< 0.046$
\\[1pt]
\hline
\hline
\label{tab:one}
\end{tabular}
\end{table*}
%

\subsection{Distortion principal component decomposition}
\label{sec:PCA}
The approximations given in the previous section are all based on simple analytical considerations. However, the CMB spectrum is given by a superposition of $\mu$, $y$ and $r$ distortions as well as the CMB monopole temperature. The situation is further complicated by the presence of large foregrounds. Also, when considering instrumental effects (number of channels, upper and lower frequency channel and frequency resolution, etc.), different spectral shapes cannot be uniquely separated and project on to each other. In this case, a principal component analysis (PCA) helps parametrizing the expected distortion shapes with a small number of distortion parameters, ranked by their expected signal to noise ratio.

Considering experimental setting similar to {\it PIXIE}, this decomposition was carried out previously \citep{Chluba2013PCA}, identifying new distortion parameters, $\mu_k$, to describe the $r$-type distortion. 
Since a minimal distortion parameter estimation will include $\mu$, $y$ and $\Delta = \Delta T/T_0$, the $r$-type distortion is defined such that none of the $\mu_k$ correlated with any of these.
For the technical details we refer to \citet{Chluba2013PCA}, but the primordial distortion signal in each frequency bin can then be decomposed as 
\bsub
\begin{align}
\label{eq:DI_vec}
\Delta I_{i}&=\Delta I_{i}^T + \Delta I_{i}^\mu+\Delta I_{i}^y+\Delta I^R_{i}
\\[1mm]
\Delta I^R_{i}&\approx \sum_k S^{(k)}_{i} \,\mu_k,
\end{align}
\esub
where $\Delta I_{i}^T$, $\Delta I_{i}^\mu$ and $\Delta I_{i}^y$ correspond to the spectral shapes of a temperature shift, $\mu$- and $y$-distortions, respectively; $\Delta I^R_{i}$ describes the $r$-type distortion, where $\mu_k$ and $S^{(k)}_{i}$ are the amplitude and distortion signal of the $k^{\rm th}$ eigenmode, respectively (see Fig.~\ref{fig:Sk}).

The signal eigenmodes, $S^{(k)}_{i}$, are associated with a set of energy release eigenvectors, $\vek{E}^{(k)}$, in discretized redshift space, which are normalized as $\vek{E}^{(k)}\cdot \vek{E}^{(l)} =\delta_{kl}$. Any given energy-release history, $\mathcal{Q}(z)=\id (Q/\rho_\gamma)/\id \ln z$, can then be written as
\beal
\label{eq:eigenmodes}
\mathcal{\vek{Q}}&\approx \sum_k \vek{E}^{(k)} \,\mu_k,
\end{align}
where $\mathcal{\vek{Q}}=(\mathcal{Q}(z_0), \mathcal{Q}(z_1), ..., \mathcal{Q}(z_{n}))^{T}$ is the energy-release vector of $\mathcal{Q}(z)$ in different redshift bins.
By construction, the eigenvectors, $\vek{E}^{(k)}$, span an ortho-normal basis, while all $\vek{S}^{(k)}$ only define an orthogonal basis (generally $\vek{S}^{(k)}\cdot \vek{S}^{(l)} \geq \delta_{kl}$). 
The mode amplitudes are then obtained as simple scalar product, $\mu_k=\vek{E}^{(k)}\cdot \mathcal{\vek{Q}}$. The eigenmodes are ranked by their signal-to-noise, such that modes with larger $k$ contribute less to the signal. In a similar way, we can write $\mu=1.401\,\vek{J}_\mu \cdot \mathcal{\vek{Q}}$, $4y=\vek{J}_y \cdot \mathcal{\vek{Q}}$ and $4\Delta T/T_0=\vek{J}_T \cdot \mathcal{\vek{Q}}$, where the $\vek{J}_i$ vectors play the role of effective visibilities, like in the integral versions, Eq.~\eqref{eq:Greens_approx_improved}. The first few energy-release eigenmodes and $\vek{E}^i$ are shown in Fig.~\ref{fig:Ek}. We note that again only heating at $z\geq 10^3$ is included in the estimates of the distortion parameters. This can underestimate the $y$-parameter at the $\simeq 10\%$ level (see Sect.~\ref{sec:results}).

Although in terms of simple scattering physics, distortion signals created by energy release at $z\lesssim \pot{{\rm few}}{5}$ should deviate from a simple $\mu$-distortion \citep{Hu1995PhD, Chluba2008c, Chluba2011therm, Khatri2012mix, Chluba2013Green}, even at lower redshift a non-vanishing additional projection on to $\mu$ is found \citep{Chluba2013PCA}, as reflected by an increase of $\vek{J}^\mu$ around $z\simeq 10^5$ (see Fig.~\ref{fig:Ek}). This enhances the recovered value of $\mu$ when the parameter estimation problem for $\mu$, $y$ and $\Delta$ is solved (see below), an effect that is mainly due to the fact that the average CMB temperature has to be determined simultaneously. 

We immediately mention that one alternative approach, which could mitigate the above problem, could be to determine the average CMB temperature using the integral properties of the $\mu$, $y$ and $r$-distortions. Since these are created by a scattering process, the number density of photons should not change. Thus, determining the effective temperature of the CMB by computing the total photon number density ($N_\gamma \propto \Tg^3$) from the measured spectrum, the distortions would not contribute under idealized assumptions. However, this procedure is not expected to work well for discretized versions of the spectra. It is furthermore complicated by the presence of foregrounds and the possibility of non-standard processes that can actually lead to non-trivial photon injection \citep{Chluba2015GreensII}. Finally, the $r$-distortion parameters would no longer remain uncorrelated, so that we do not explore this avenue any further.

\subsection{Results from the different methods}
\label{sec:results}
We are now in the position to explicitly compute the $\mu$- and $y$-parameters for the different distortion scenarios discussed above. With the PCA, we are furthermore able to obtain the eigenmode amplitudes, $\mu_1$ and $\mu_2$. We stop at the second residual distortion eigenmode, since observing $\mu_2$ is already very futuristic for standard scenarios. We also mention that the values for $y$ are only used as a comparison, since the $y$-distortion from the low-redshift Universe is much larger in all cases. 

\begin{figure}
\centering 
\includegraphics[width=\columnwidth]{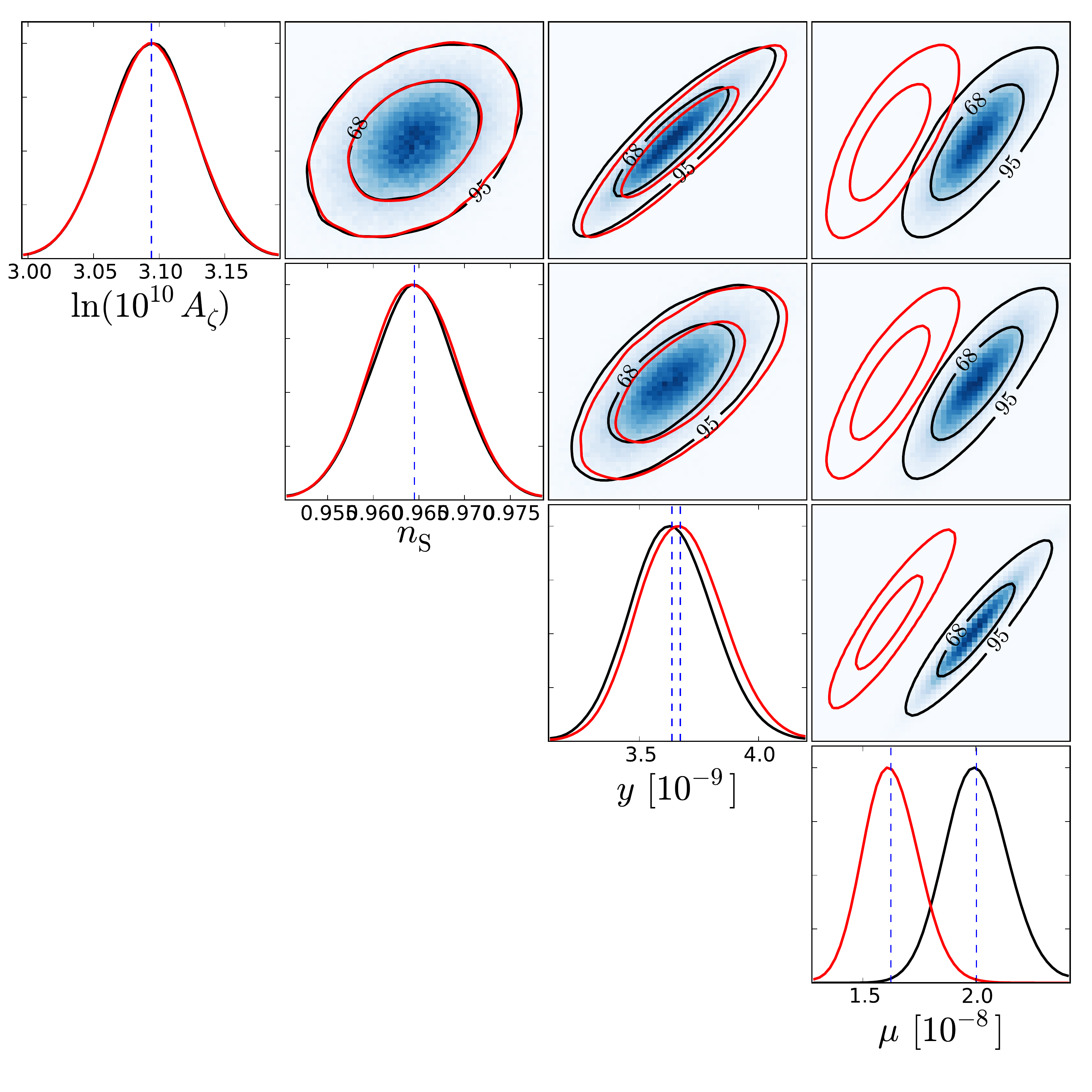}
\caption{Comparison of the posterior distributions for the dissipation scenario I (Table~\ref{tab:one}) obtained with method B (red contours) and the PCA (black contours). The vertical lines indicate the mean values. Method B predicts lower values for $\mu$ than the distortion eigenmode analysis.}
\label{fig:M2_PCA}
\end{figure}

In our estimates, we include the measurement uncertainties for the relevant $\Lambda$CDM parameters \citep{Planck2015params}. For the dissipation scenarios, these are mainly related to the power spectrum parameters, while for the adiabatic cooling distortion it is the baryon density (assuming standard BBN). The results are summarized in Table~\ref{tab:one}. For the dissipation scenarios, we obtained the error estimates by using the relevant covariance matrix for the {\it Planck} data, while the error for the adiabatic cooling effect was directly estimated using Gaussian error propagation.

\begin{figure}
\centering 
\includegraphics[width=\columnwidth]{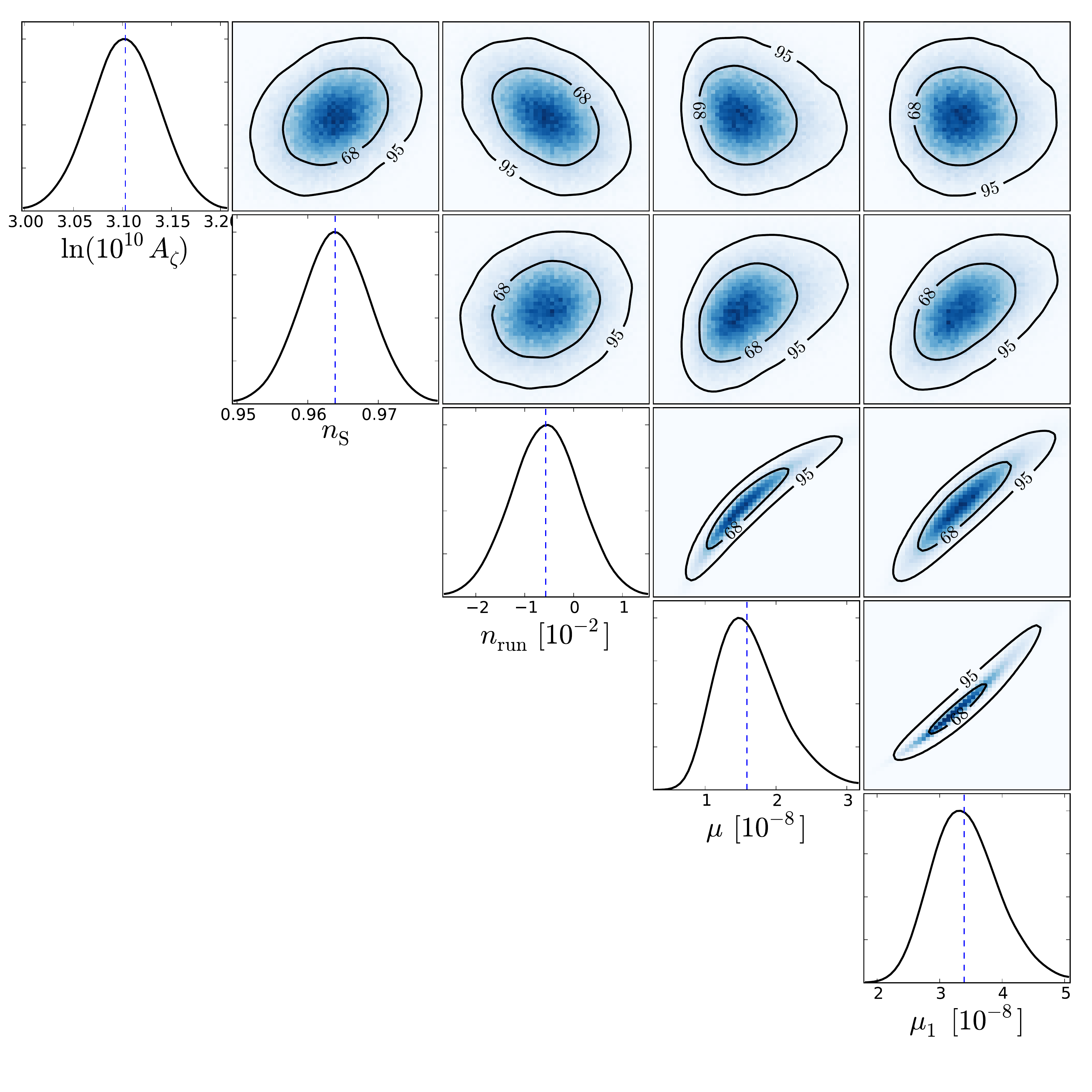}
\caption{Posterior distributions for the dissipation scenario II (Table~\ref{tab:one}) obtained with the PCA. We omitted $y$ as its posterior remains fairly Gaussain. The vertical lines indicate the mean values.}
\label{fig:DII_PCA}
\end{figure}

\begin{table}
\centering
\caption{Explicit projections of the full {\tt CosmoTherm} output using the distortion eigenmodes for {\it PIXIE}-like settings. The last column also gives the estimates $1\sigma$ error for {\it PIXIE} in its current design \citep{Chluba2013PCA}, which degrades quickly for the $\mu_k$. In parenthesis, we show estimates for the expected significance in terms of distortion measurements.}
\begin{tabular}{c cc c}
\hline
\hline
Parameter &  Dissipation I & Adiabatic cooling & {\it PIXIE} $1\sigma$
\\
\hline
$ y/10^{-9}$ & $3.54 \;(\simeq 3.0\sigma)$ & $-0.623 \;(\simeq 0.5\sigma)$ & $1.20$ 
\\[1pt]
$ \mu/10^{-8}$ & $2.00\;(\simeq 1.5\sigma)$  & $-0.334\;(\simeq 0.2\sigma)$  &  $1.37$   
\\
$ \mu_1/10^{-8}$ & $3.82\;(\simeq 0.3\sigma)$ & $\,\,\,-0.588\;(\simeq 0.04\sigma)$ & $14.8$  
\\[1pt]
$ \mu_2/10^{-9}$ & $\!\!\!\!-1.18\;(\simeq 0.0\sigma)$ & $-0.054\;(\simeq 0.0\sigma)$ & $761$  
\\
\hline
\hline
\label{tab:two}
\end{tabular}
\end{table}

For the $y$-parameter estimates, methods A and B are equivalent and give results which are quite close to those of the PCA, which should be considered the most precise representation of what would be recovered in a distortion analysis. The methods C and D are also equivalent, but overestimate the $y$-parameter by $\simeq 5\%-10\%$ in comparison to the PCA. The recovered error bars for all methods are very comparable.
For the $\mu$-parameter, all methods are slightly different. The PCA always gives $20\%-30\%$ larger values. The best agreement with the PCA is achieved by methods A and C. Again all methods give very similar estimates for the expected errors.

In Fig.~\ref{fig:M2_PCA}, we highlight the differences in the predicted $y$ and $\mu$-parameters for the dissipation scenario I obtained with method B and the PCA\footnote{The figure was obtained using the Markov Chain Monte Carlo (MCMC) tool of the {\tt Greens} software package \citep{Chluba2013Green} available at {\tt www.Chluba.de/CosmoTherm}.}. The errors are dominated by uncertainties in the power spectrum parameters. The result for $y$ agrees quite well, while the result for $\mu$ is biased low by $\simeq 2.6\,\sigma$ with method B, a difference that needs to be taken into account when interpreting future distortion measurements. The result from method B is very close to what was recently discussed in \citet{Cabass2016} for $\mu$.

In Fig.~\ref{fig:DII_PCA}, we show the posterior for dissipation scenario II (see Table~\ref{tab:one}). This no longer is a $\Lambda$CDM case, as for the standard inflation model running is negligible. However, it illustrates how well {\it Planck} data might constrain running and we will return to this case below when forecasting spectral distortion constraints.
While in the case without running the posteriors for the distortion parameters remained fairly Gaussian, when including running those for $\mu$, $\mu_1$ and $\mu_2$ become non-Gaussian. Since the {\it Planck} data prefers slightly negative running, implying less power at small scales, the median for $\mu$ decreases. Due to the long leaver arm, small changes in $\nrun$ have a large effect on $\mu$ and $\mu_1$, so that their uncertainties increase significantly relative to the case without running (see Table~\ref{tab:one}). Conversely, this means that distortion measurements have constraining power in particular for $\nrun$ \citep[e.g.,][]{Chluba2012}. Again, the expected recovered value for $\mu$ is slightly larger than what \citet{Cabass2016} find, whose central value is more close to that of methods A-C. However, in this case the significance of the bias is only $\simeq 1\sigma$ due to increased uncertainty in the prediction.

\begin{figure*}
\centering 
\includegraphics[width=\columnwidth]{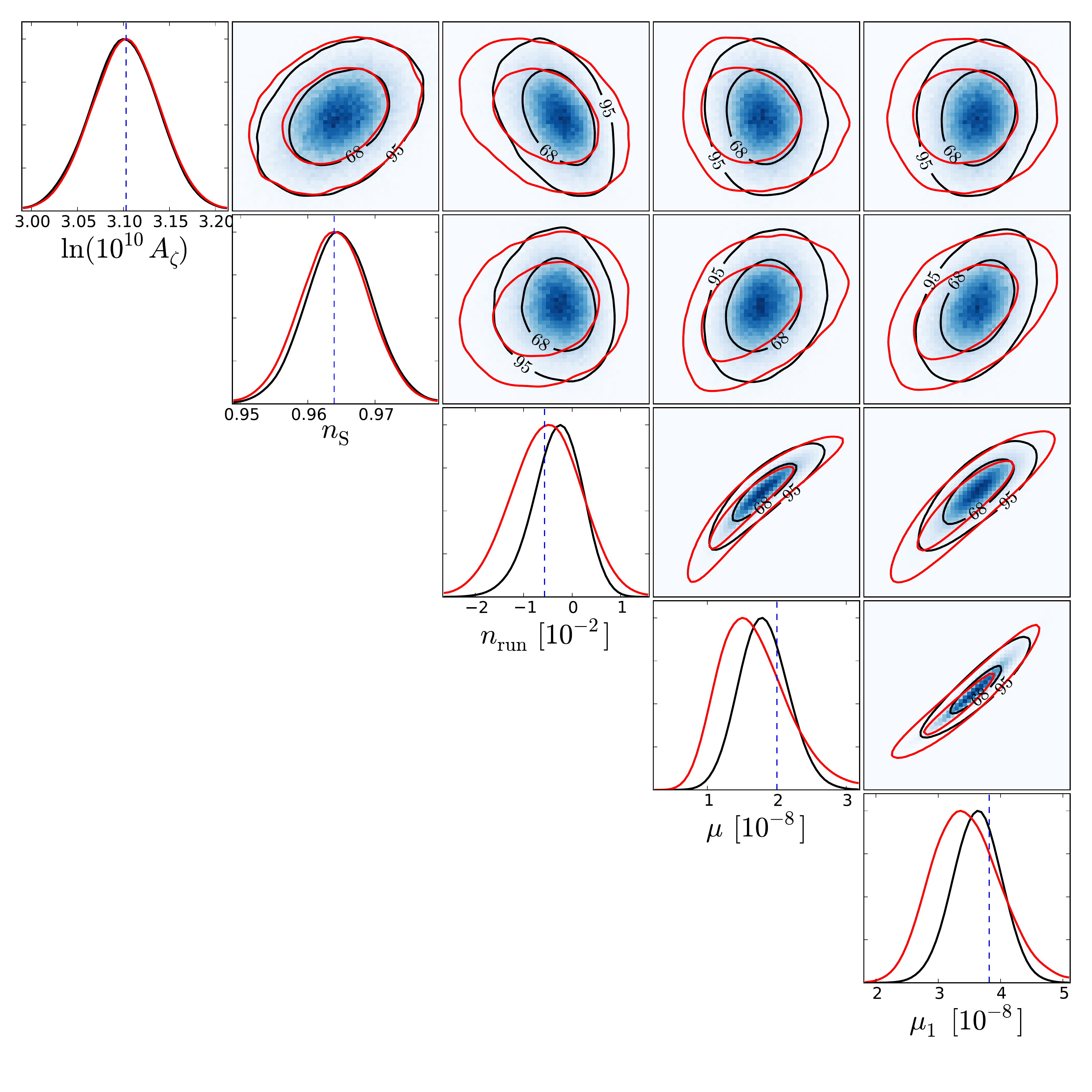}
\hspace{7mm}
\includegraphics[width=\columnwidth]{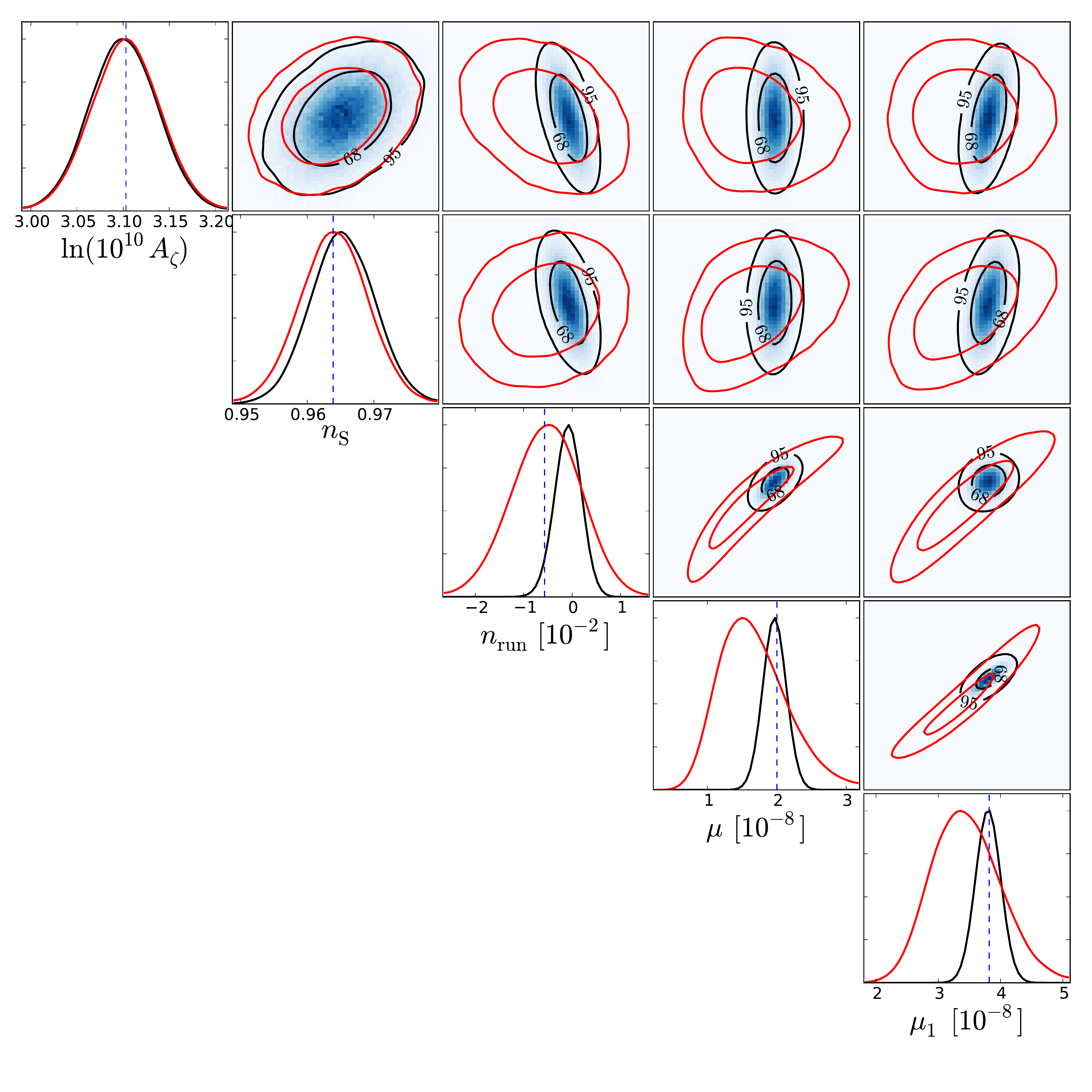}
\caption{Posteriors for different combinations of data sets. In both panels, the red lines indicate the {\it Planck} TT,TE,EE+lowP+{\it PIXIE} ($\equiv$ basically like without {\it PIXIE}) constraints for the extended model with running. The black contours show the {\it Planck}+$3.4\times${\it PIXIE} (left panel) and  {\it Planck}+$10\times${\it PIXIE} (right panel) constraints. Vertical lines indicate the fiducial values for each data set. Adding spectral distortions helps diminishing the uncertainty in the values of $\nrun$ by a factor of $\simeq 3$ for {\it Planck}+$10\times${\it PIXIE}.}
\label{fig:SD_const}
\end{figure*}

\vspace{-3mm}
\subsection{Distortion parameters for full {\tt CosmoTherm} outputs}
\label{sec:COSMOTherm}
We mention that by explicitly using the output from {\tt CosmoTherm} for the dissipation scenario I and the adiabatic cooling case we find values summarized in Table~\ref{tab:two}. We can see that the agreement is excellent in comparison with the Green's function projections that were used for the results presented in Table~\ref{tab:one}. The differences are mainly noticeable for the $y$-parameters, but even there they are $\lesssim 10\%$. The main reason for the difference is because with {\tt CosmoTherm}  we include the adiabatic cooling effect all the way to redshift $z=200$, while in the Green's function approximation we stop at $z=10^3$. This gives another $y\approx -\pot{0.1}{-9}$ correction, which one can simply add by hand. It is possible to improve this in the code, but since this is not distinguishable from the much larger $y$-distortion created at lower redshifts, we neglect it. We mention that we include part of the cosmology-dependence of the Green's function, but at low frequencies this is only approximate.

\begin{table*}
\centering
\caption{Improvement of constraints on the small-scale power spectrum by combining {\it Planck} with a {\it PIXIE}-like experiment for different channel sensitivities. For the spectral distortion parameters, we also show the effective significance of the signal with respect to the spectral distortion measurement. The distortion amplitude $\mu_2$ remained undetectable ($\lesssim 0.02\sigma$) with distortions alone and thus remains a {\it derived} parameter even for $10\times${\it PIXIE} sensitivity. In the last column we show the {\it Planck} $\Lambda$CDM values for comparison.}
\begin{tabular}{c cccc c}
\hline
\hline
Parameter & {\it Planck} alone &  +{\it PIXIE} &  +$3.4\times${\it PIXIE} &  +$10\times${\it PIXIE} & {\it Planck} $\Lambda$CDM values
\\
\hline
$\ln(10^{10} A_{\rm s})$
& $3.103 ^{+ 0.036 } _{- 0.036 }$ 
& $3.103 ^{+ 0.037 } _{- 0.037 }$
& $3.101 ^{+ 0.037 } _{- 0.037 }$
& $3.100 ^{+ 0.036 } _{- 0.036 }$
& $3.094 ^{+ 0.034 } _{- 0.034 }$
\\[2pt]
$\nS$
& $0.9639 ^{+ 0.0050 } _{- 0.0050 }$ 
& $0.9640 ^{+ 0.0050} _{- 0.0050 }$
& $0.9647 ^{+ 0.0049 } _{- 0.0048 }$
& $0.9653 ^{+ 0.0048 } _{- 0.0047 }$
& $0.9645 ^{+ 0.0049 } _{- 0.0049 }$
\\[2pt]
$10^3\nrun$
& $\!\!\!\!-5.7^{+ 7.1 } _{- 7.1 }$ 
& $\!\!\!\!-5.2 ^{+ 6.9 } _{- 7.2}$
& $\!\!\!\!-2.8 ^{+ 4.6 } _{- 5.1 } $
& $\!\!\!\!-0.81 ^{+ 2.4 } _{- 2.5 }  $
& $0 $
\\
\hline
$ \mu/10^{-8}$ 
& $1.59 ^{+ 0.54 } _{- 0.40 }$
& $1.62 ^{+ 0.55 } _{- 0.42 }\;(1.2\sigma)$  
& $1.81 ^{+ 0.36 } _{- 0.33 }\;(4.5\sigma)$ 
& $1.993 ^{+ 0.053 } _{- 0.053 }\;(15\sigma)$ 
& $2.00 ^{+ 0.14 } _{- 0.13 }$ 
\\[2pt]
$ \mu_1/10^{-8}$ 
& $3.39 ^{+ 0.58 } _{- 0.49 }$
& $3.43 ^{+ 0.58 } _{- 0.52 }\;(0.23\sigma)$ 
& $3.63 ^{+ 0.38 } _{- 0.38 }\;(0.83\sigma)$
& $3.819 ^{+ 0.044 } _{- 0.044 }\;(2.6\sigma)$
& $3.81 ^{+ 0.22 } _{- 0.20 }$
\\[2pt]
$ \mu_2/10^{-9}$ 
& $\!\!\!\!-2.79 ^{+ 2.05 } _{- 1.53 }$
& $\!\!\!\!-2.69 ^{+ 2.08 } _{- 1.61 }\;(0\sigma)$ 
& $\!\!\!\!-2.02 ^{+ 1.42 } _{- 1.31 }\;(0\sigma)$ 
& $\!\!\!\!-1.28 ^{+ 0.43 } _{- 0.43 }\;(0\sigma)$ 
& $\!\!\!\!-1.19 ^{+ 0.22 } _{- 0.20 }$ 
\\
\hline
\hline
\label{tab:SD_const}
\end{tabular}
\end{table*}

\vspace{-3mm}
\section{Forecast for {\it PIXIE}-like experiments}
\label{sec:forecasts}
We now discuss the prospects for detecting the different distortion signals. 
First of all, the $y$-parameter contributions are all insignificant compared to the $y$-parameter caused by low-redshift processes (Sect.~\ref{sect:reion}). Although in terms of sensitivity, the contributions summarized in Table~\ref{tab:one} are significant at the level of a few $\sigma$ for {\it PIXIE} in its current design, these cannot be separated and are thus not discussed further \citep[see][]{Chluba2013fore, Chluba2013PCA}.  

The $\mu$-distortion amplitude for the dissipation scenario I is detectable at the level of $\simeq 1.45\sigma$ \citep[see also][]{Chluba2013PCA}. This is a factor of $\simeq 3.4$ short of a clear $\simeq 5\sigma$ detection of the standard $\Lambda$CDM signal \citep[cf.,][]{Chluba2012}. One simple improvement in the sensitivity is achieved by halving the number of channels. For a Fourier Transform Spectrometer (FTS), such as in {\it PIXIE}, this improves the detection limits by a factor of $\simeq \sqrt{2}$, since about twice as much time can be spent on each sample and but the total number of collected photons remains the same\footnote{The effective channel sensitivity improves by a factor of $2$, but half the number of channels are available to constrain the distortions so that overall only $\simeq \sqrt{2}$ is gained \citep[compare][]{Kogut2011PIXIE}.}. However, this also reduces the number of frequency channels in the CMB regime and thus degrades our ability to reject foregrounds, requiring a more detailed optimization. Similarly, changes in the distribution of channels could improve the performance of {\it PIXIE}, but also call for a more careful analysis of foreground effects. Another way to improve the sensitivity is by extending the mission duration or by changing the time spent on spectral distortions versus $B$-mode science \citep{Kogut2011PIXIE, Chluba2013fore}. Thus, factors of a few improvement in the raw spectral sensitivity seem to be within reach of current technology, although ultimate limitations by foregrounds need to be addressed carefully (Sect.~\ref{sec:foregrounds}).

One interesting questions is at what effective sensitivity can a spectral distortion measurement confirm $\Lambda$CDM assuming the simplest slow-roll inflation model \citep[see][for review]{Baumann2009}, which predicts $\nrun\simeq -\frac{1}{2}(\nS-1)^2 \simeq \pot{6}{-4}$ \citep{Starobinsky1980}. To answer this, we assume a fiducial model for the small-scale power spectrum as in dissipation scenario I, which gives $\mu=\pot{2.0}{-8}$. However, we then compute the posterior assuming {\it Planck} constraints for the case with running. Adding spectral distortions should then pull the best-fitting solution closer to $\nrun\simeq 0$. We also add a large $y$-distortion signal\footnote{The uncertainty in the value of $y$ from low redshifts removes about 2/3 of the constraining power of to the dissipation signal. Without this large `foreground' one would achieve a factor of $\simeq 1.6$ improvement of the constraint on $\nrun$ with {\it PIXIE} in its default setup already.} $y=\pot{2}{-6}$ and a small shift in the CMB monopole temperature. Both parameters are estimated simultaneously with the power spectrum parameters using the MCMC tool of the {\tt Greens} software package.

In Fig.~\ref{fig:SD_const}, we illustrate this aspect for two channel sensitivities. Derived combined parameter constraints are summarized in Table~\ref{tab:SD_const}. 
For the fiducial sensitivity of {\it PIXIE}, adding spectral distortions does not improve the constraints on power spectrum parameters significantly. 
For $\simeq 3.4$ times higher channel sensitivity, we find an $\simeq 4.5\sigma$ detection of $\mu$ in terms of the uncertainty of the distortion measurement and a factor of $\simeq 1.4-1.5$ improvement of the error on $\nrun$, which starts to move towards the $\Lambda$CDM value. The uncertainties in the values of $A_{\rm s}$ and $\nS$ are not improved much (see Fig.~\ref{fig:SD_const}).
At $10$ times the sensitivity of {\it PIXIE}, which is similar to the spectrometer of the {\it PRISM} concept \citep{PRISM2013WPII}, the constraints are further tightened by adding spectral distortions, reducing the error on $\nrun$ by a factor of $\simeq 3$ over {\it Planck} alone. We also find an $\simeq 15\sigma$ detection of the $\mu$-distortion parameter and a marginal $\simeq2.6\sigma$ detection of $\mu_1$ from distortions alone in this case. This sensitivity would furthermore make the CRR detectable at the level of a few $\sigma$ \citep{Vince2015}.

In the near future, ground-based observations (Stage-IV CMB) in combination with the upcoming spectroscopic galaxy surveys, eBOSS and DESI, aim at improving limits on $\nrun$ by factors of a few \citep{Abazajian2015inflation}. One other target is to obtain improved constraints on neutrino physics \citep{Abazajian2015}. As the $\mu$-distortion from the dissipation scenario is mostly sensitive to $\nrun$, adding spectral distortion measurements could help improving these limits by alleviating some existing degeneracies. However, enhanced versions of {\it PIXIE} are required in this case.

We highlight that the small-scale damping process is the dominant source of $\mu$-distortions in $\Lambda$CDM. This implies that any significant departure from the expected signal (Table~\ref{tab:two}) inevitably points towards new physics. If the signal is much lower, then the small-scale CMB power spectrum, around wavenumbers $k\simeq 740\,\Mpc^{-1}$, where most of the $\mu$-signal is created \citep[see][for illustration]{Razi2015}, either has a much lower amplitude \citep{Chluba2012, Chluba2012inflaton}, or an enhanced cooling process through the coupling of another non-relativistic particle to the CMB is required \citep[e.g.,][]{Yacine2015DM}. Conversely, if the $\mu$-distortion signal is much larger than expected, then the small-scale power spectrum could be strongly enhanced, possibly containing a localized feature \citep{Chluba2012inflaton, Chluba2015IJMPD}, or another heating mechanism \citep[e.g., a decaying particle][]{Hu1993b, Chluba2011therm, Chluba2013fore, Dimastrogiovanni2015} has to be at work. Thus, spectral distortions provide a powerful new avenue for testing $\Lambda$CDM cosmology without purely relying on an extrapolation from large ($k\lesssim 1\,\Mpc^{-1}$) to small scales ($1\,\Mpc^{-1}\lesssim k\lesssim \pot{\rm few}{4}\,\Mpc^{-1}$).

\vspace{-3mm}
\subsection{Importance of refined foreground modeling}
\label{sec:foregrounds}
It is clear that for the success of spectral distortion measurements, the name of the game will be foregrounds. The biggest challenge is that, aside from the large $y$-distortion introduced at late times, all known foregrounds are orders of magnitudes larger than the primordial signals. This means that tiny effects related to the spectral and spatial variation of the foreground signals need to be taken into account. Ways to tackle this problem are i) to measure the spectrum in as many individual channels as possible, ideally with high angular resolution and sensitivity, and ii) to exploit synergies with other future or existing datasets to inform the modeling of averaged signals. In both cases, refined modeling of the foregrounds with extended parametrizations are required to capture the effects of averaging of spatially varying components across the sky.

An FTS concept like {\it PIXIE} pushes us into a qualitatively different regime in terms of its spectral capabilities, where instead of playing with a few channels we have a few hundred at our disposal. Most of these channels are at high frequencies ($\nu\gtrsim 1\,{\rm THz}$), above the CMB bands and can be used to subtract the dust and cosmic infrared background components at lower frequencies \citep{Kogut2011PIXIE}. Simple, commonly used two-temperature modified blackbody spectra \citep[e.g.,][]{Finkbeiner1999} will not provide sufficient freedom to capture all relevant properties of the averaged dust spectrum.
Existing maps from {\it Planck} \citep[e.g.,][]{Planck2013components, Planck2015components} can be used to estimate the effect of spatial variations of the dust temperature across the sky. Similar to the superposition of blackbodies with varying temperature (Sect.~\ref{sec:sup}), this will cause new spectral shapes because of the i) {\it beam average} and ii) {\it all-sky averaging}, which need to be captured by extended foreground parametrizations (Chluba et al., 2016, in preparation). The associated parameters have to be directly determined in the distortion measurement, as existing data are not expected to provide sufficient information down to the noise level of experiments like {\it PIXIE}, but can be used to guide the modeling and parametrization.

Similar comments apply to the modeling of the synchrotron and free-free components at low frequencies. Albeit spectrally quite smooth, the superposition of spatially varying power-law spectra, $I_\nu\simeq A_0 \, (\nu/\nu_0)^{\alpha}$ [where $I_\nu$ denotes the intensity], causes new spectral shapes that depend on {\it moments} of the underlying distribution functions. The common addition of curvature to the spectral index, $I_\nu\simeq A_0 \, (\nu/\nu_0)^{\alpha+\frac{1}{2}\beta \ln(\nu/\nu_0)}$ \citep[e.g., compare][]{Remazeilles2015}, is too restrictive. Extending the list of curvature parameters,
\beal
\label{eq:power_sum_exp_power}
I^{\rm p}_{\nu}&\approx A_0 \, (\nu/\nu_0)^{\alpha+\frac{1}{2}\beta \ln(\nu/\nu_0)+\frac{1}{6}\gamma \ln^2(\nu/\nu_0)+\frac{1}{24}\delta \ln^3(\nu/\nu_0)+\ldots},
\end{align}
can be shown to capture all the new degrees of freedom for the superposition of power-law spectra, although this generalization does not have the best convergence properties (Chluba et al., 2016, in preparation). In Eq.~\eqref{eq:power_sum_exp_power}, the coefficients $\alpha$, $\beta$, $\gamma$ and $\delta$ are directly related to the aforementioned moments of the spectral index distribution functions. 
These have to be determined with the distortion measurement, while external data sets \citep[e.g., C-BASS,][]{Irfan2015} can likely only be used to guide the modeling at the level of sensitivity targeted by future CMB spectroscopy.

Additional important foreground components are due to anomalous microwave emission \citep{Draine1998, Yacine2009, Planck2011AME}, various narrow molecular lines (e.g., CO, HCN, HCO, etc.), zodiacal light \citep{Planck2014zodi} and the integrated flux of CO emission throughout cosmic history \citep{Righi2008b, Mashian2016}, all of which will also affect future CMB imaging experiments. In all cases, one will have to make use of properties of the underlying physical mechanism to motivate refined foreground parametrizations. 
The primordial distortion signals caused by energy release are all {\it unpolarized}, which is another important property to exploit. Systematic effects related to the absolute calibration could furthermore be tested using the motion-induced leakage of monopole signals into the CMB dipole \citep{Balashev2015}. It was also pointed out that the relativistic correction signal (see Sect.~\ref{sect:reion}) could cause significant confusion for the $r$-distortion signals \citep{Hill2015}, which has to be modeled more carefully. All the above challenges need to be addressed to demonstrate the full potential and feasibility of future spectroscopic CMB experiments, a problem we are currently investigating. 
For additional recent discussion of the foreground separation problem for the CMB monopole see \citet{Mayuri2015}, \citet{Vince2015} and \citet{deZotti2015}. 

\vspace{-3mm}
\section{Conclusions}
Within $\Lambda$CDM, a range of guaranteed distortion signals is expected. Here, we summarized these distortions and also provided an assessment of the expected uncertainties in their prediction (Sect.~\ref{sec:distortions_LCDM} and Fig.~\ref{fig:signals}). The largest signals is due to the formation of structures and reionization process at late times (Sect.~\ref{sect:reion}). The signal is characterized by a $y$-type spectrum, with a $y$-parameter reaching $\simeq \pot{\rm few}{-6}$. Although the uncertainty in the amplitude of this contribution is rather large, this signal will be detectable with a {\it PIXIE}-like experiment at very high significance, even allowing us to determine the small relativistic temperature correction caused by abundant low-mass haloes with temperature $\Te \simeq 10^6\,\Kel$ \citep[see][]{Hill2015}. This promises a way to solve the missing baryon problem and constrain the first sources of reionization.

The largest primordial signal expected within $\Lambda$CDM is due to the damping of small-scale acoustic modes (Sect.~\ref{sec:damp}). The $y$-distortion contribution is swamped by the low-redshift signal, but the $\mu$-distortion should be detectable with a slightly improved version of {\it PIXIE}. Using {\it Planck} data, we find $\mu\approx\pot{(2.00\pm0.14)}{-8}$ for the standard $\Lambda$CDM cosmology, where the uncertainty is dominated by those of power spectrum parameters. This value includes the fact that temperature shift, $\mu$, $y$ and $r$-distortions are not uniquely separable, but that these parameters have to be determined simultaneously. As our comparison of different approximation methods shows (Sect.~\ref{sec:estimates}), simple estimates for $\mu$, based solely on scattering physics arguments, are expected to underestimate the recovered/measured value obtained with future distortion experiments by $\simeq 20\%-30\%$ (see Table~\ref{tab:one}).

Our simple forecasts show (Sect.~\ref{sec:forecasts}), that by combining CMB spectral distortion measurements with existing {\it Planck} data, one can achieve an $\simeq 40\%-50\%$ improvement of the error on $\nrun$ for $\simeq 3.4$ times the sensitivity of {\it PIXIE}. At this sensitivity, also an $\simeq 5\sigma$ detection of $\mu$ from distortion measurements alone can be expected. About $\simeq 10$ times the sensitivity of {\it PIXIE} is required for a marginal $\simeq2.6\sigma$ detection of the first $r$-distortion parameter, $\mu_1\approx \pot{(3.82 \pm 0.22)}{-8}$, assuming $\Lambda$CDM. This sensitivity would furthermore render the CRR detectable at the level of a few $\sigma$ \citep{Vince2015} and deliver a factor of $\simeq 3$ improvement of the error on $\nrun$ (Fig.~\ref{fig:SD_const} and Table~\ref{tab:SD_const}).
A combination of CMB spectral distortion measurements with upcoming ground-based experiments (e.g., Stage-IV CMB) could help further tightening constraints on power spectrum parameters and neutrino physics; however, enhanced versions of {\it PIXIE} are required to achieve this.

In Sect.~\ref{sec:foregrounds}, we gave a brief discussion of the spectral distortion foreground challenge. Simple, physically-motivated extensions of the foreground parametrizations are required to capture the effect of averaging of spatially varying spectral components inside the instrumental beam and across the sky. While existing data can be used to inform the underlying models, at the high sensitivity targeted by future spectroscopic missions, these foreground parameters ultimately have to be determined in the measurements. FTS concepts provide many hundred channels, which should allow us to extend the parameter list of refined foreground models from a few to tens, promising a path towards detailed spectral distortion measurements that we will investigate in the future.

We close by mentioning that in spite of all the successes of $\Lambda$CDM, there are open puzzles, such as the nature of dark matter and dark energy, to name the obvious ones. Spectral distortions are sensitive to {\it new physics} and any departure from the expected $\Lambda$CDM predictions for the high-$z$ signals will inevitably point in this direction. For example, if at early times dark matter coupled to baryons or photons, then this will leave an effect on the CMB spectrum \citep[e.g.,][]{Yacine2015DM}, potentially diminishing the net distortion.
Also, significantly higher or significantly lower power at small scales, responsible for the $\mu$-distortion signal ($k\simeq 740\,\Mpc^{-1}$), are necessarily related to departures from simple slow-roll inflation \citep{Chluba2012inflaton, Chluba2013PCA, Clesse2014, Chluba2015IJMPD, Cabass2016}. The presence of unaccounted relicts (e.g., gravitinos or moduli) or excited states of dark matter, decaying early enough to leave the CMB anisotropies unaffected, could furthermore play a role \citep{Hu1993b, Chluba2011therm, Chluba2013fore, Chluba2013PCA, Dimastrogiovanni2015}. 

This illustrates only a few of the interesting new directions that CMB spectral distortions measurements could shed light on, and the big challenge will be to disentangle different effects to allow us draw clear conclusions. We can only look forward to the advent of real distortion data in the future.

\small
\section*{Acknowledgements}
JC cordially thanks Nick Battaglia, Giovanni Cabass, Colin Hill, Alessandro Melchiorri and Enrico Pajer for valuable feedback on the manuscript, and Steven Gratton for helpful advice on {\it Planck} data. 
JC is supported by the Royal Society as a Royal Society University Research Fellow at the University of Manchester, UK. 

\small 

\bibliographystyle{mn2e}
\bibliography{Lit}

\end{document}

%% file: Befehle.tex
\newcommand{\eV}{{\rm eV}}
\newcommand{\keV}{{\rm keV}}

\newcommand{\Kel}{{\rm K}}

\newcommand{\cm}{{\rm cm}}

\newcommand{\Mpc}{{\rm Mpc}}
\newcommand{\GHz}{{\rm GHz}}

\newcommand{\expf}[1]{{{\rm e}^{#1}}}

\newcommand{\nS}{n_{\rm S}}

\newcommand{\nrun}{n_{\rm run}}

\newcommand{\kD}{k_{\rm D}}

\newcommand{\id}{{\,\rm d}}

\newcommand{\beq}{\begin{equation}}   %

\newcommand{\eeq}{\end{equation}}   %

\newcommand{\beqa}{\begin{eqnarray}}   %

\newcommand{\eeqa}{\end{eqnarray}}   %

\newcommand{\beal}{\begin{align}}
\newcommand{\enal}{\end{align}}

\newcommand{\bspl}{\begin{split}}

\newcommand{\espl}{\end{split}}

\newcommand{\bsub}{\begin{subequations}}

\newcommand{\esub}{\end{subequations}}

\newcommand{\bmulti}{\begin{multline}}   %

\newcommand{\beqm}{\begin{mathletters}}   %

\newcommand{\eeqm}{\end{mathletters}}   %

\newcommand{\Ne}{N_{\rm e}}

\newcommand{\Te}{T_{\rm e}}

\newcommand{\Tg}{T_{\gamma}}

\newcommand{\vek} [1]{\mbox{\boldmath${#1}$\unboldmath}}

\newcommand{\pot}[2]{#1 \times 10^{#2}}

\newcommand{\Yp}{Y_{\rm p}}
